\documentstyle[aps,prl,preprint,epsfig]{revtex}    
\tighten 
\newcommand{\ket}[1]{\,| \, {#1} \,\rangle  \,}
\newcommand{\bra}[1]{\,\langle \, {#1} \,|  \,}
\newcommand{\tgh} {{\rm tgh} \,} 
\newcommand{\bq}{\begin{eqnarray}}
\newcommand{\eq}{\end{eqnarray}}

\begin{document}  
\title{Decoherence in Josephson qubits}
\author{G.~Falci$^{(1)}$, 
E.~Paladino$^{(1)}$, 
and R.~Fazio$^{(2)}$}
\address{
$^{(1)}$ Dipartimento di Metodologie Fisiche e 
	Chimiche (DMFCI), Universit\`a di Catania,
        viale A. Doria 6, 95125 Catania, Italy \& 
Istituto Nazionale per la Fisica della Materia, UdR Catania. \\
$^{(2)}$Scuola Normale Superiore, 56126 Pisa, Italy \& NEST-INFM, Pisa.}


\maketitle

\begin{abstract}
A high degree of quantum coherence is a crucial 
requirement for the implementation of quantum logic devices.
Solid state nanodevices seem particularly
promising from the point of view of
integrability and flexibility in the design. 
However decoherence is a serious limitation, due to
the presence of many types low energy excitations in the 
``internal'' environment and of ``external'' sources due to the control
circuitery. 
Here we study both kind of dephasing in a special implementation, 
the charge Josephson qubit, however many of our results are applicable
to a large class of solid state qubits. This is the case of $1/f$ noise 
for which we introduce and study a model of an environment of bistable
fluctuatiors. External sources of noise are analized in terms of
a suitable harmonic oscillator environment and the explicit mapping on 
the spin boson model is presented.
We perform a detailed investigation of 
various computation procedures (single shot measurements, repeated 
measurements) and discuss the problem of the information needed to
characterize the effect of the environment.
For a fluctuator environment with $1/f$ spectrum memory effects turn out to
be important. 
Although in general information beyond the power spectrum is needed, 
in many situations this results in the knowledge of only one more 
microscopic parameter of the environment. This allows to determine which 
degrees of freedom of the environment are effective sources of decoherence 
in each different physical situation considered.
\end{abstract}


\section{Introduction}
Solid state coherent system are at the forefront of present day research
because of the perception that large scale integration 
may be combined with new physical properties to yield new paradigms for
nanoelectronics.
A concrete example is quantum 
computation~\cite{kn:qcomp,kn:nielsen,kn:clark-ed}
with solid state devices, with several 
proposals~\cite{kn:loss,kn:rmp,kn:teor,makhlin99,kn:nature} and few recent  
experiments~\cite{kn:Nakamura1,kn:vion,kn:two-qubit}. 
More generally the ability of controlling the dynamics of a complex quantum
system would open a wide scenario for both fundamental and applied physics.

Controlled dynamics may be achieved if first it is possible to prepare (write)
and measure (read) a set of $N$ observables $\{Q_i\}$~\cite{kn:nielsen}. 
This set defines the basis of the so called 
{\em computational states} $\ket{\{q_i\}}$. If moreover it is possible to
tune the Hamiltonian of the system, then the dynamics of a generic state
\begin{equation}
\label{eq:state}
\ket{\psi, t} = \sum_{q_1 \dots q_N} c_{q_1 \dots q_N}(t) 
\; \ket{{q_1, \dots, q_N}}
\end{equation}
may be controlled.
In general $\ket{\psi, t}$ is a {\em superposition} of computational 
states and moreover if $c_{\{q_i\}}(t)$ cannot be factorized in individual
functions $c_{q_i}(t)$, then $\ket{\psi, t}$ is 
{\em entangled}~\cite{kn:qcomp,kn:nielsen}.
Quantum algorithms use superpositions to produce constructive interference 
towards the correct answer and entanglement for the speed up of information 
processing. 
Thus certain calculations which are practically impossible on a classical 
computer could be performed if {\em coherence} could be preserved.
Coherence usually denotes situations when a well defined relation 
between the components $c_{\{q_i\}}(t)$ of the state Eq.(\ref{eq:state}) 
exists. In our case this simply means that the nanodevice can be described 
by the pure state Eq.(\ref{eq:state}), the relation between the 
$c_{\{q_i\}}(t)$ being the solution of the Schr\"odinger equation. 
Loss of coherence is due to the fact that the Hilbert space of the device is
much larger than the computational space~\cite{kn:zurek,kn:palma96}. 
The {\em system} (defined by the set $\{Q_i\}$) interacts with the 
{\em environment} (defined by all the other observables needed to complete 
the set)~\cite{kn:weiss}. Even a weakly coupled environment may cause 
{\em decoherence}~\cite{kn:zurek}, i.e. it may destroy the 
phase relation between $c_{\{q_i\}}(t)$. The system should be described
by a density matrix $\rho(t)$ rather than a pure state as 
Eq.(\ref{eq:state}).

Considering a larger Hilbert space is needed because 
the nanodevice is a many-body object and $\{Q_i\}$ is only a small
set of collective variables. Decoherence from these ``internal'' sources
represents a serious limitation due to the presence of many low energy 
excitations in the solid state environment. Clever protocols
and technological progress may reduce these effects, but even in 
an idealized situation there is an ``external'' environment of apparata
(for preparation, measurement, tuning of the Hamiltonian during time 
evolution) which are themselves quantum systems enlarging the overall 
Hilbert space. 
To appreciate the importance of the external environment, 
notice that of course we would be happy with a system we can manipulate 
at will, but to obtain easy tunability we have to open a port 
to the external world and this determines decoherence. 

\begin{figure}
\centerline{\resizebox{75mm}{35mm}{
\includegraphics{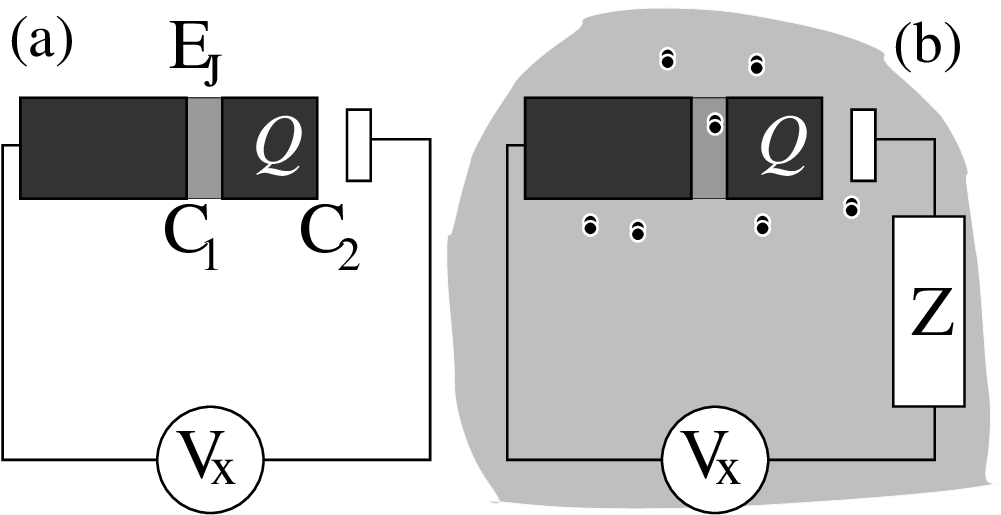}}
\hfill
\resizebox{!}{35mm}{\includegraphics{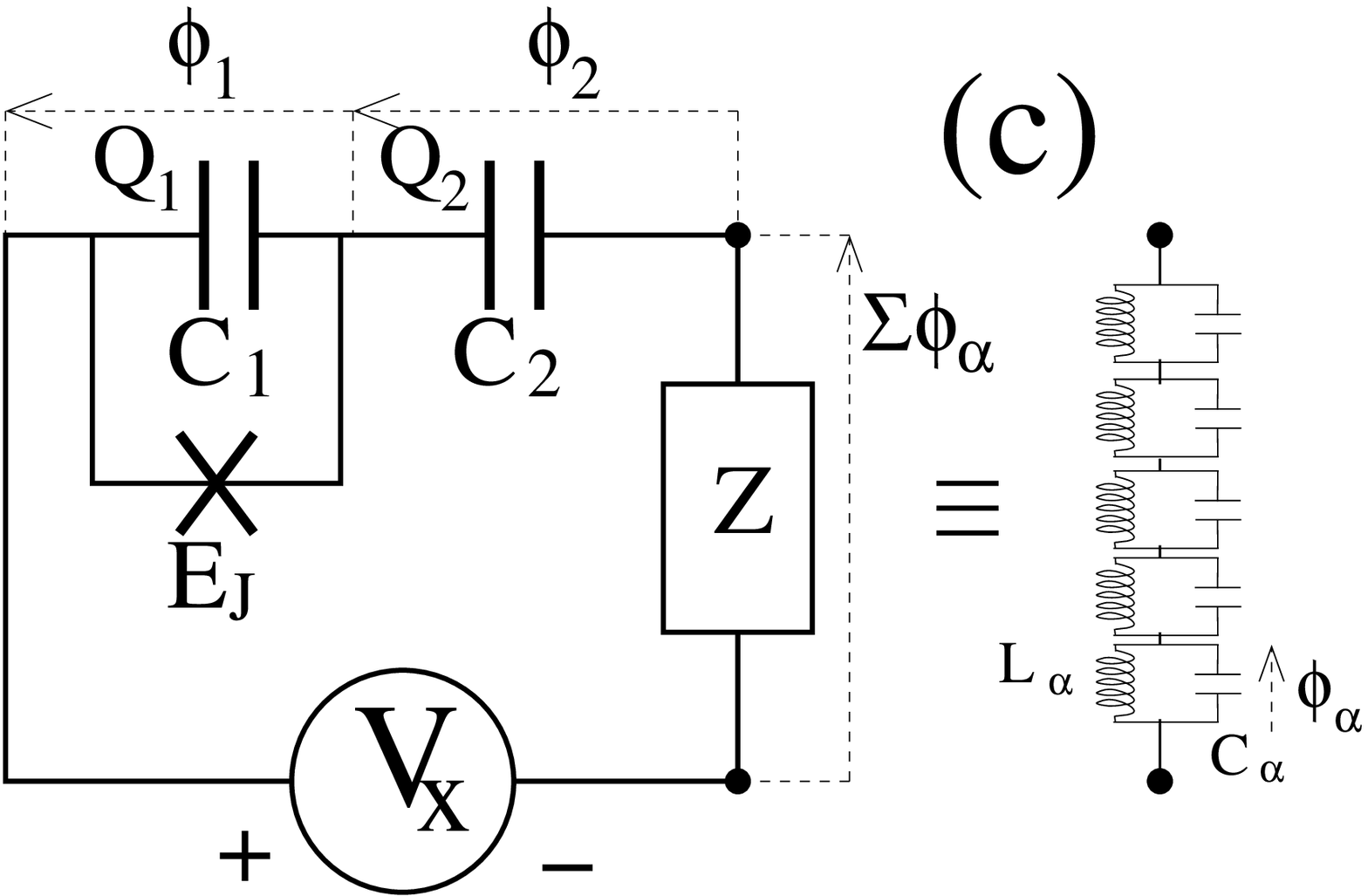}}}
\caption{(a) Ideal charge-Josephson qubit formed by a superconducting
single electron box; (b) Charge Josephson qubit in the presence of
the circuit and the substrate, with sources of electromagnetic and 
offset charge fluctuactions;
(c) Equivalent circuit for the qubit in the electromagnetic environment: 
the impedance $Z(\omega)$ is modeled by a suitable infinite $LC$ 
transmission line.}
\label{fig:cjqubit}
\end{figure}
In this work we consider the charge-Josephson qubit~\cite{makhlin99}. 
It is a superconducting island, where the charge $Q$ can be stored, 
connected to a 
circuit via a Josephson junction and a capacitance 
$C_2$ (Fig.\ref{fig:cjqubit}a). 
Adding a charge polarizes the surroundings and costs energy, provided by 
the voltage source $V_x$. Then $V_x$  may be used to fix $Q$ to the desired 
value. On the other hand Josephson tunneling mixes the charge states. 
Under suitable conditions 
(charging energy $E_C=  e^2/2(C_1+C_2)$ much larger than 
the Josephson coupling $E_{J}$ and temperatures $k_B T \ll E_J$ )
only two charge states are important and 
the system realizes a qubit with Hamiltonian
\begin{equation}
\label{eq:qubit-hamiltonian}
{\cal H}_Q \,=\,  \, \frac{\varepsilon}{2} \, \sigma_z  -
 \frac{E_J}{2} \,\sigma_x \qquad ; \qquad 
\varepsilon(V_x) =4 E_{C} (1-C_2V_x/e)
\end{equation}
where the eigenstates $\{\ket{n}: n=0,1\}$ of $\sigma_z$ represent a 
well defined extra number $n$ of Cooper pairs in the island and span the
computational space.
Superconducting qubits~\cite{kn:rmp,kn:teor,kn:nature} are 
important because they are the
only solid state implementations where coherence in a single qubit has been 
observed in the time domain~\cite{kn:Nakamura1,kn:vion} and promising 
experiments exist for two-qubit systems~\cite{kn:two-qubit}. 
Problems on decoherence can thus be posed in a realistic perspective. 

We will 
study models of environments which describe 
decoherence due to fluctuations of the external circuit 
and to $1/f$ noise produced by charges which may be trapped close to 
the device (Fig.\ref{fig:cjqubit}b). 
These have been recognized to be major sources of errors 
in charge-Josephson qubits, because 
the computational states $\{\ket{n}\}$ are coupled to charges 
moving in the 
environment\footnote{We will not discuss here other applications of our work, 
but we stress that 
charge noise is ubiquitous in the solid state and it is important also 
for solid state qubits based on the electron spin, where proposed 
implementations of two-qubits gates use the Coulomb interaction. Moreover the 
models we consider can be directly applied to flux noise in flux-Josephson 
qubits~\cite{kn:mooji,kn:clark}.}. 
The system plus environment Hamiltonian reads 
\begin{equation}
\label{eq:syst-env-hamiltonian}
{\cal H} \,=\, {\cal H}_Q - \, \frac{1}{2} \, \hat{E} \,\sigma_z  + {\cal H}_E
\end{equation}
where ${\cal H}_E$ describes the environment, coupled to the system via the
operator $\hat{E}$, which acts as an extra contribution to the polarization 
$\varepsilon$.

Since we are interested on measurements on the qubit, the
standard road-map is to calculate the reduced density matrix 
$\rho(t) = \mathrm{Tr}_E \{W(t)\}$ where $W(t)$ is the system plus environment
density matrix. Equations for $\rho(t)$ can be written in terms of statistical 
information on the environment, i.e. correlation 
functions like $\langle \hat{E}(t_1)\hat{E}(t_2) \dots \rangle_E$ and 
system-environment correlations, which makes the problem formidable.
In practice we hope that in order to make predictions we do 
not really need such a detailed knowledge of the microscopic parameters 
which define ${\cal H}_E$. This may happen in two remarkable cases, namely 
if the environment is weakly coupled and fast, and if the environment 
is modeled by harmonic oscillators. Then 
all the information needed on the environment is encoded in the power 
spectrum of the coupling operator $\hat{E}$
\begin{equation}
\label{eq:power-spectrum}
S(\omega) \,=\, \int_{-\infty}^{\infty} \hskip-6pt dt \;\frac{1}{2} \;
\langle \hat{E}(t)\hat{E}(0) + \hat{E}(0)\hat{E}(t)  \rangle 
\; \mathrm{e}^{i \omega t}
\end{equation}
Roughly speaking if the environment is weakly coupled and fast on the time
scales typical of decoherence, then the system is unable to probe it in great
detail. If the environment is made of harmonic oscillators all its 
equilibrium statistical properties can be derived from the power 
spectrum\footnote{The Caldeira Leggett model has been proposed in this 
spirit to describe the environment in Macroscopic Quantum Tunneling and 
Coherence in Josephson circuits~\cite{kn:leggett}}.
Unfortunately this is not enough to describe all the effects of a 
solid state environment, since low-energy excitations may 
determine {\em memory effects}  and 
in general specific gates are sensitive to different details of the 
environment.

To conclude this introduction we notice that although decoherence comes from 
the entanglement of the system with its environment, the reduction of the 
amplitude of the coherent signal in specific experiments 
may often be studied in less fundamental terms. For instance one 
may think to the environment as producing a classical stochastic 
field which couples to the qubit. In this way spontaneous emission is missed
but, apart from that, this may be a useful point of view 
for practical estimates of the effect of solid state environments.

\section{Simple estimates of decoherence}
\subsection{Master Equation}
\begin{figure}
\begin{center}
\resizebox{!}{35mm}{\includegraphics{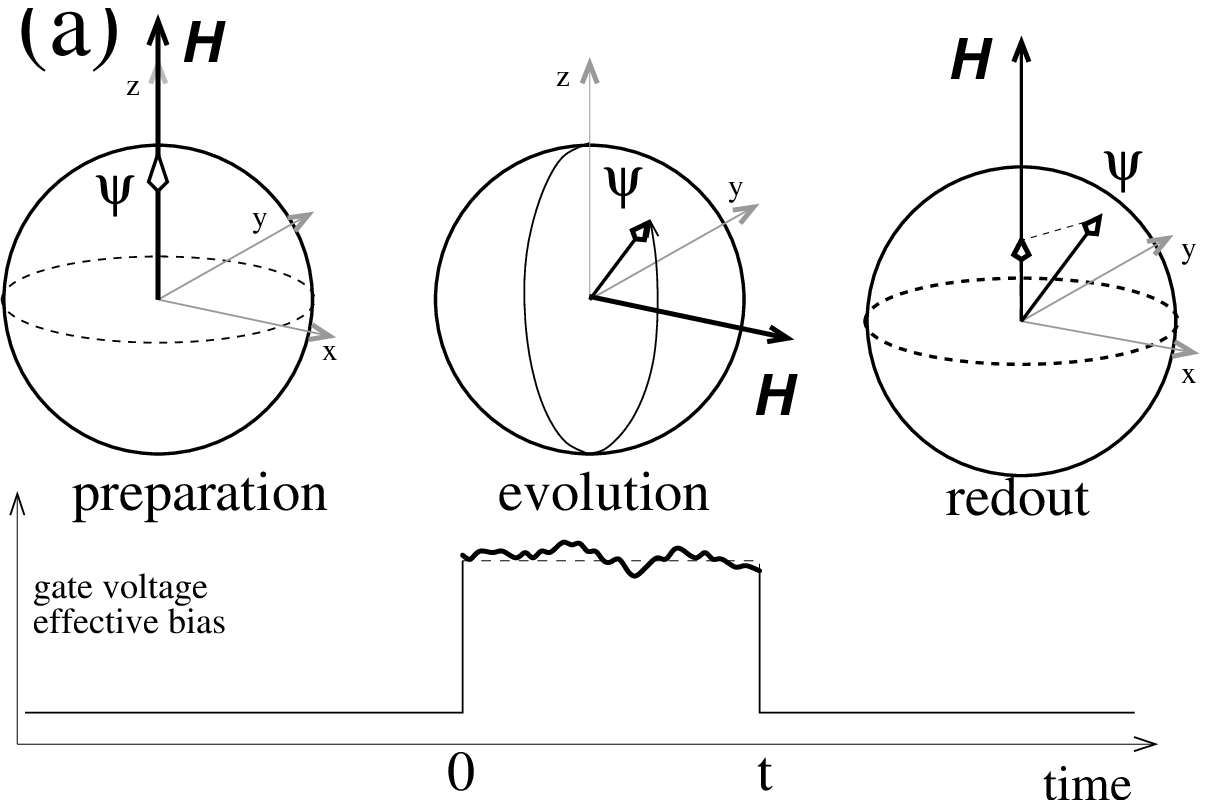}}
\hspace{15mm}
\resizebox{!}{35mm}{\includegraphics{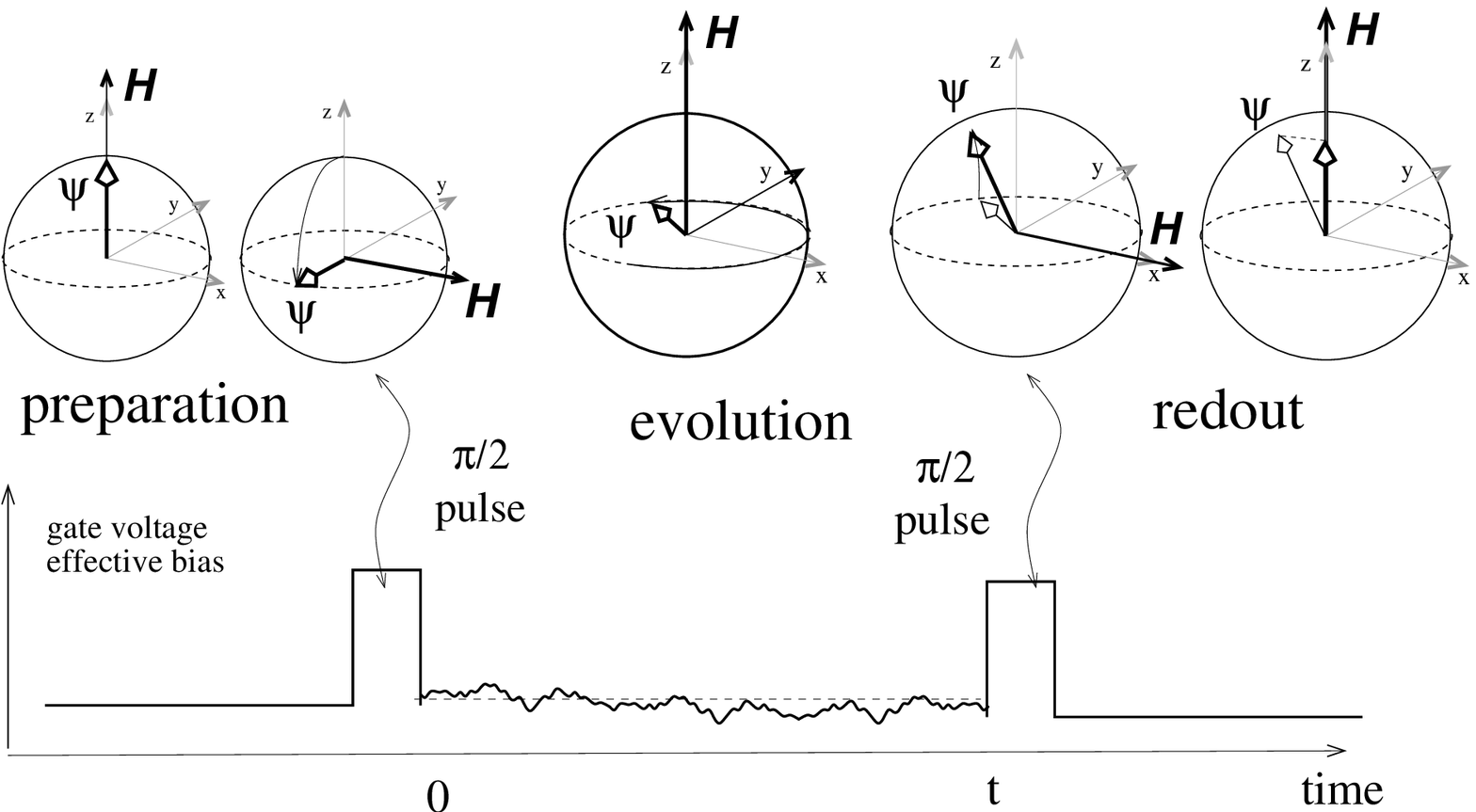}}
\end{center}
\caption{Bloch sphere representation of the typical protocol
used to generate a phase shift in a qubit, by operating with a slightly
noisy gate 
(a) at the optimal operation point, $\theta=\pi /2$,
and (b) at $\theta=0$, when ``pure'' dephasing occurs. Notice that for 
charge-Josephson qubit charge states can be easily set and measured, so 
preparation and redout occur along the $\hat{z}$ axis.}
\label{fig:single-shot}
\end{figure}

For a weakly coupled environment the trace on the environmental 
degrees of freedom can be approximately calculated in several ways, 
in second order in the interaction~\cite{kn:lax-redfield}. 
A standard approach
leads to the  Markovian Master equation in the basis of
the eigenstates of ${\cal H}_{Q}$~\cite{kn:cohen} 
\begin{equation}
\label{eq:master}
\dot{\rho}_{ij} \;=\; - i \omega_{ij} \,\rho_{ij} + 
\sum_{mn} 
\; {\cal R}_{ijmn} \; \rho_{mn}
\end{equation}
where $\omega_{ij}$ is the difference of the energy eigenvalues.
The relaxation tensor ${\cal R}_{ijmn}$ depends on combinations of 
quantities of the kind
$\int_0^{\infty} \hskip-2pt dt \; {\cal C}^{ 
\raisebox{5pt}{\tiny $>$} \hskip-5pt \raisebox{-0pt}{\tiny $<$}
}_{ijkl}(t)$ 
and may be calculated if the Green's function of the
environment
$
i \, G^{ >}(t) = \langle \hat{E}(t) \hat{E}(0)\rangle_{E}$, or its 
Fourier transform 
$2 \, S(\omega)/( 1 + \mathrm{e}^{- \beta \omega}) 
$, is known 
\footnote{The correlators ${\cal C}^{ 
\raisebox{5pt}{\tiny $>$} \hskip-5pt \raisebox{-0pt}{\tiny $<$}
}_{ijkl}(t)$ 
read
${\cal C}^{\raisebox{5pt}{\tiny $>$}}_{ijkl}(t) = Tr_E \left \{ 
w_E(0) \langle i | e^{i H_E t} H_{int}  e^{-i H_E t} | j \rangle
\langle l | H_{int} | k \rangle
\right \}$ and
${\cal C}^{\raisebox{5pt}{\tiny $<$}}_{ijkl}(t) = Tr_E \left \{ 
w_E(0)\langle i | H_{int} | j \rangle
 \langle l | e^{i H_E t} H_{int}  e^{-i H_E t} | k \rangle
\right \}$, for a factorized initial density matrix 
$W(0)= w_E(0) \otimes \rho(0)$.
The system-environment interaction term
for the Hamiltonian Eq.(\ref{eq:syst-env-hamiltonian}) reads
$H_{int}=- \hat{E} \sigma_z /2$.}.
In deriving this equation it has been assumed 
that the environment is always at 
equilibrium and it is fast, i.e. $G^{ >}(t)$ should decay on a time scale 
$\tau_c$ which is the smallest scale in the problem. Often the
sum in Eq.(\ref{eq:master}) can be restricted to the secular terms
(terms such that $\omega_{ij} = \omega_{mn}$) and the result for 
the relaxation rate $\Gamma_R$, governing the exponential decay of 
the populations $\rho_{ii}$ towards equilibrium, and for the decoherence
rate $\Gamma_\phi$ which describes the vanishing of the coherences 
$\rho_{ij}$, is readily found
\begin{equation}\label{eq:master-result}
{\Gamma_R}
= \frac{1}{2} \, \sin^2 \theta \; S(\Omega)
\rule[-20pt]{0pt}{6pt}
\qquad ; \qquad 
\Gamma_\phi
= \Gamma_\phi^{0} + \frac{1}{2} \, \Gamma_R =
\frac{1}{2} \, \cos^2 \theta \; S(0)
+ \frac{1}{2} \, \Gamma_R
\end{equation}
where $\Omega = \sqrt{\epsilon^2+E_J^2}$ is the bare level splitting and
$\mathrm{tan} \, \theta = - E_J/\varepsilon$ is the angle characterizing 
${\cal H}_Q$ in the Bloch sphere representation.
As we will see the major problems 
in solid state come from the so called adiabatic term 
$\Gamma_\phi^{0}$, which depends on low frequencies of the environment. 
The result Eqs.(\ref{eq:master-result}) suggests that an optimal operation 
point is $\theta = \pi/2$ and that dephasing only depends on the power 
spectrum of the environment at the operating frequency $\Omega=E_J$.
A typical protocol is depicted in Fig.\ref{fig:single-shot}a.

\subsection{Pure dephasing}
For $\theta = 0$ the system is sensitive 
to low frequencies of the environment. 
This  case, usually referred to as ``pure dephasing'', is special 
in that 
the Hamiltonian~(\ref{eq:syst-env-hamiltonian}) commutes with $\sigma_z$. 
The charge in the island is conserved and no relaxation occurs. However if
we prepare the qubit in a superposition of charge states the system will 
dephase and consequently the coherences will decay 
(a typical protocol is depicted in Fig.\ref{fig:single-shot}b). 

If 
the initial density matrix is factorized, $W(0)= w_E(0) \otimes \rho(0)$, 
it is possible to write an exact expression for the coherences only in terms
of the environment~\cite{kn:palma96,kn:PRL}.
\begin{equation}
\label{eq:gamma-pure-dephasin-general}
\rho_{01}(t) = \rho_{01}(0) \; \mathrm{e}^{-\Gamma(t) - i \delta E(t)} \qquad ; \qquad
\Gamma(t) = - \ln \left|
\mathrm{Tr}_E \bigl\{ w_E(0) \; 
\mathrm{e}^{i{\cal H}_{-1}t} \, \mathrm{e}^{-i{\cal H}_{1}t}
\bigr\} \right|
\end{equation}
where ${\cal H}_\eta = {\cal H}_{E} - (\eta/2) \,\hat{E}$.

\subsection{Model for circuit fluctuations}
In order to apply the above results we now specify a model for the environment
which describes fluctuations of the external circuit, modeled by an external 
impedance $Z(\omega)$ (Fig.\ref{fig:cjqubit}b). On the classical level the
effect is accounted for by substituting $V_x \to V_x + X(t)$ 
in Eq.(\ref{eq:qubit-hamiltonian}), where $X(t)$ are the classical voltage
fluctuations $X(t)$ at the impedance.
The phenomenological quantum model is obtained by applying the following rules.
First we substitute $V_x \to V_x + X$ in  Eq.(\ref{eq:qubit-hamiltonian}). 
Then we introduce a set of harmonic oscillators to describe 
the impedance and 
to weight quantum fluctuations of $X$
$$
{\cal H}_{E} = \sum_\alpha 
\left\{
{p_\alpha^2 \over 2 m_\alpha} +  {m_\alpha \omega_\alpha^2 \over 2} \,  
x_\alpha^2 \right\} 
\qquad ; \qquad X = \sum_\alpha c_\alpha x_\alpha
$$
which specifies the Hamiltonian 
(\ref{eq:syst-env-hamiltonian}). Finally the power spectrum of $X$ is identified by the voltage fluctuations at 
$Z(\omega)$ 
$$ 
S_X(\omega) \,=\,
|\omega| \; {\cal R}e \, { Z(\omega) \over 1 + i \omega Z(\omega) 
C_{eff}} \;
\mathrm{ctgh} { \beta |\omega| \over 2} \,=\,
J(\omega) \; \mathrm{ctgh} { \beta |\omega| \over 2}
$$
where $C_{eff}=C_1 C_2/(C_1+C_2)$ and to make contact with the standard
notation~\cite{kn:weiss} we introduced the 
spectral 
density of the environment $J(\omega)$. This allows to identify 
$S(\omega) = (4 E_C C_2/e)^2 S_X(\omega)$ and to estimate the rates using 
Eq.(\ref{eq:master-result}). At the optimal point the quality factor 
$ E_J/\Gamma_\phi =
 \left( C_1/C_{eff} \right)^{2}\, 2 R_Q / (\pi R) \sim 10^6$ 
for typical
values of the parameters (we took
$Z(\omega)=R$ and $R_Q=h/(4e^2)$ is the superconducting quantum of 
resistence), 
which would allow a single qubit gate to 
work perfectly. Adiabatic dephasing would lead to an even larger 
$E_J/\Gamma_\phi$, by a factor $E_J/K_B T$. 

\subsection{$1/f$ noise} 
Numerous experiments have shown that the performance of single electron 
tunneling (SET) devices strongly suffer from fluctuations of 
background charges (BC) 
located in the vicinity of the junctions~\cite{zorin96,nakamura-echo}. 
Their behavior can be visualized as a random extra polarization $E(t)$ 
which produces voltage fluctuations at the device. 
Voltage noise can be measured in SET transistors, and it 
has been observed to have the $1/f$ form up to $1 \, kHz$~\cite{zorin96}. 
Using experimental data we may identify 
$S(\omega) =  16 \pi {\cal A} E_C^2/ \omega$ 
where $\cal A$ is about $10^{-6}$~\cite{zorin96}
and Eq.(\ref{eq:master-result}) at the optimal point leads
to a quality factor $E_J / \Gamma_\phi \sim 10^3$.
This procedure underlies
two questionable assumptions: first, 
one should extrapolate observed $1/f$ noise up to frequencies 
$E_J \sim 10 \, GHz$, second, one should make the ad hoc assumption that 
there are nonequilibrium impurity charges at such frequencies, much larger 
than the
temperature\footnote{Measurements on the accuracy of single-electron traps
showed indirect effects of $1/f$ noise at high frequencies~\cite{kn:noise-set}.They were 
interpreted within a similar ``one photon'' approach wich requires the
ad hoc assumption that there are nonequilibrium impurity charges at such 
frequencies. This analysis yields the better value 
${\cal A} \sim 10^{-8}$ but is certainly non conclusive.
}.
Apart from this Eqs.(\ref{eq:master-result}) badly fails 
in estimating $\Gamma_\phi^0$ which is proportional to 
the inverse of the small
(and unmeasurable) low-frequency cut-off $\gamma_m$ of $1/f$ noise. 
In order to have a more reliable estimate it has been proposed 
that one may describe 
the environment as a set of harmonic oscillators~\cite{kn:cottet}. 
In this case the exact
expression Eq.(\ref{eq:gamma-pure-dephasin-general}) 
may be evaluated~\cite{kn:palma96}, yielding
\begin{equation}
\label{eq:gamma-pure-dephasing-oscillators}
\Gamma_{osc}(t)
\;=\; \int_{0}^{\infty} \hskip-2pt {d\omega \over \pi} \;
S(\omega) \;
{1 - \cos \omega t \over \omega^2} \,. 
\end{equation}
For times $t \ll 1/ \gamma_M$, where $\gamma_M$ is the high-frequency cut-off 
of the $1/f$ noise,  $\Gamma_{osc}(t)$ is approximated by
$\Gamma_{osc}(t) \approx 8 {\cal A} E_C^2 
\ln\left({\gamma_m \over \gamma_M}\right)t^2$.
The result
is still dependent on $\gamma_m$ in a relevant way: each decade of noise, 
including noise produced by very slow fluctuators, 
gives exactly the same contribution to dephasing. 
It has been proposed the recipe to use instead
an ad hoc effective low-frequency cut-off which gives resonable
results, but in order to have a clear picture of what is going 
on one should take more seriously the discrete character 
of charge noise. In particular the main difficulty in treating 
the $1/f$ environment by the above standard methods is that the large 
majority of degrees of 
freedom are much slower than all time scales of the evolution of the 
system, so they give rise to important {\em memory effects}.
As a consequence different gates are sensitive to different details of the 
environment.

\section{Model for $1/f$ noise}
\label{sec:model and methods} 
In this section we introduce a simple model of an environment 
which yields $1/f$ noise, which is a suitable set of bistable 
fluctuators~\cite{kn:PRL}. 
We first consider the fluctuators to be sources of a classical stochastic 
process, i.e. each fluctuator switches between two configurations with 
total rate $\gamma_i$ and determines an extra polarization of the qubit 
given by $\frac{v_i}{2} p_i(t)$, where $p_i(t)=\pm 1$. 
The total extra polarization, 
$E(t) = \sum_i \frac{v_i}{2} p_i(t)$, 
has power spectrum
$
S(\omega)= \sum_i (v_i/2)^2 \hskip-2pt
\int\limits_{-\infty}^{\infty} \hskip-6pt dt 
\, (\overline{p_i(t)p_i(0)} - \overline{\delta p}^2)
\, \mathrm{e}^{i \omega t}  = 
\sum_i v_i^2/2 
(1- \overline{p}^2) \, \gamma_i/(\gamma_i^2 + \omega^2)
$, 
where the overline means average on the stochastic processes. 
The standard assumption~\cite{kn:weissman} of a distribution of switching rates 
$P(\gamma) \propto 1/ \gamma$ for $\gamma \in [\gamma_m,\gamma_M]$ and 
zero elsewhere leads to $1/f$ noise, 
$S(\omega)= 
\{\pi (1- \overline{p}^2)\,  n_d \,\overline{v^2} /(4 \,\, \ln \, 10)\} \,\,
\omega^{-1}
$
for frequencies  
$\omega \in [\gamma_m,\gamma_M]$  ($n_d$ is the number of
fluctuators per noise decade).
Already at this level it is clear that the environment presents a large
number of slow fluctuators, so memory effects are important.

We introduce a quantum model~\cite{kn:PRL,bauernschmitt93} by describing 
each fluctuator as a 
localized impurity
level connected to a band~\cite{kn:mahan}. The system plus environment 
Hamiltonian reads
\begin{eqnarray}
	{\cal H}
	&=& {\cal H}_Q \,-\, {1 \over 2} \, \sigma_z \, 
	\sum_i v_i  \,  b_i^{\dagger} b_i \, 
	\, + \, \sum_i 	{\cal H}_i
\label{eq:hamiltonian-backcharges}\\ 
{\cal H}_i &=& \varepsilon_{c i} b_i^{\dagger} b_i +
		\sum_{k} [T_{ki}  c_{ki}^{\dagger} b_i + \mbox{h.c.} 
]
		 +  \sum_{k} \varepsilon_{ki}  c_{ki}^{\dagger} c_{ki} \, .
\nonumber
\end{eqnarray}
Here ${\cal H}_i$ describes an isolated BC: 
the operators $b_i$ ($b^{\dagger}_i$) destroy (create) an electron 
in the localized  level $\varepsilon_{ci}$.
This electron may tunnel, with amplitude $T_{ki}$ to a band
described by the operators $c_{ki}$, $c^{\dagger}_{ki}$ and the energies 
$\varepsilon_{ki}$. For simplicity  
we assume that each localized level is connected to a distinct band.
An important scale is the total switching rate 
$\displaystyle{\gamma_i=2 \pi {\cal N}(\epsilon_{ci})|T_i|^2}$
(${\cal N}$ is the density of states of the electronic band, 
$|T_{ki}|^2 \approx |T_i|^2$), which characterizes the classical relaxation 
regime of each BC.
Finally the coupling with the qubit is such that each BC  
produces a bistable extra bias $v_i$. 

It may be argued that the impurity model introduced here describes impurities 
in metals, rather than in insulating 
substrates. However our aim is to discuss consequences of the discrete nature
of the BCs and this is the simplest model embedding this feature. 
Moreover this model has been introduced and used in 
Ref.~\cite{bauernschmitt93} to
successfully explain experiments on charge trapping in systems of
small tunnel junctions very similar to the charge-qubit. 

Our aim is to investigate the effect of the BC 
environment on the dynamics of the qubit.
The picture which emerges from our analysis is that the contribution of 
the single BC in dephasing the qubit depends on the ratio 
$g_i \equiv v_i/\gamma_i$. 
As a consequence, we distinguish between 
two different kinds of BC: the ones with $v/\gamma \ll 1$, which we 
call {\em weakly coupled} and the ones in the other regime, which we 
call {\em strongly coupled}. Concerning these latter notice that we are 
interested to a
physical situation where the $v_i$ are so small that the energy scale 
associated to 
the total extra bias produced by the set of BCs is much smaller than $\Omega$,
so strongly coupled charges means small $\gamma_i$. 

Strongly coupled BCs give rise to memory effects and moreover differences 
of their statistical properties from those
of an oscillator environment (cumulants higher than the second are 
nonvanishing) may be relevant~\cite{kn:weissman}. As a consequence the results 
Eqs.(\ref{eq:master-result})
 may be inapplicable. One possibility is to calculate higher 
orders in the Master equation, but even if one assumes that the environment is
still modelled by harmonic oscillators, the corrections for $1/f$ spectrum are 
of the same order of the leading terms~\cite{kn:shnirman}. 
Here we follow a different strategy, namely we enlarge the ``system'' which
allows to treat more accurately the dynamics of the BCs, at all orders in the 
couplings $v_j$. In the modified roadmap we consider {\em only} the 
continuos electron bands in Eq.(\ref{eq:hamiltonian-backcharges}) as the 
environment and investigate the reduced dynamics of the system composed 
by the qubit and the BCs.
We find that weakly coupled charges behaves as a source of gaussian noise, whose effect is fully characterized by the power spectrum $S(\omega)$.  On the other hand, from our quantum mechanical treatment it emerges that the decoherence due to strongly coupled charges shows pronunced features of their discrete character.
In the following we will specialize our analysis to the two cases $\theta = 0, \pi/2$.

\section{Qubit at the optimal point}
\begin{figure}
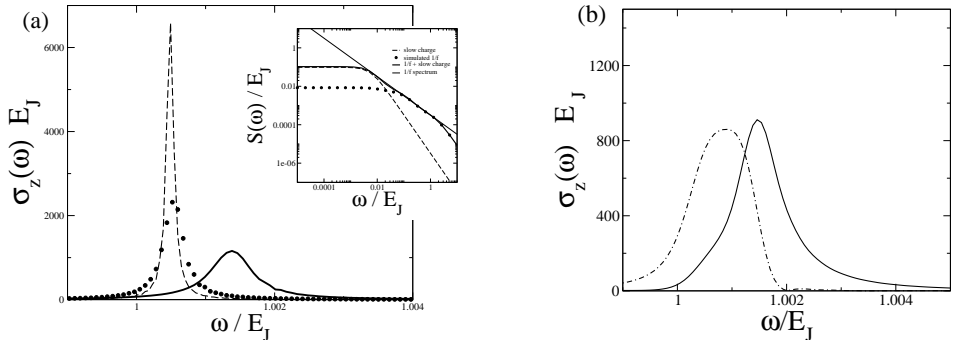

\centerline{\resizebox{!}{45mm}{
\includegraphics{degenerazione1.eps}}
\hspace{-25mm}\raisebox{48pt}{\resizebox{!}{25mm}{\includegraphics{degenerazione2.eps}}} 
\hspace{10mm}
\resizebox{!}{45mm}{\includegraphics{fig2mod.eps}}}
\caption{(a) The Fourier transform $\sigma_z(\omega)$ 
for a set of weakly coupled BCs 
plus a single strongly coupled BC (solid line). The separate effect of 
the coupled slow BC alone ($g_0= 8.3$, dashed line) and of the 
set of weakly coupled BCs 
(dotted line), is shown for comparison.
In the inset the corresponding power spectra: notice that at $\omega = E_J$
the power spectrum of the extra charge alone (dashed line) is very small. 
In all cases the noise level at $E_J$ is fixed to the value 
${\mathrm S}(E_J) / E_J \approx 3.18 \times 10^{-4}$.
(b)
The Fourier transform $\sigma_z(\omega)$ 
for a set of weakly coupled BCs plus a strongly coupled BC  
($v_0 / \gamma_0 =61.25$) prepared in the ground (dash-dotted line) or in the 
excited state (solid line). 
}
\label{figure:equil}
\end{figure}

If the environment is made of a single BC we can implement the new roadmap 
by solving the Master equation Eq.(\ref{eq:master}) for a system composed 
by the qubit and the BC. This method is unpractical for a large number 
$N$ of BCs since 
the number of equations becomes $2^{2N}-1$. So  
we study the general problem by using the Heisenberg equations of motion. For 
the average values of the qubit observables
$\langle \sigma_{\alpha} \rangle$, $\alpha=x,y,z$ we obtain ($\hbar=1$)
\begin{equation}\label{eq:heisenberg} \quad
\langle \dot{\sigma}_x \rangle = 
          \sum_i^N v_i \, \langle b_i^{\dagger} b_i \sigma_y \rangle
  \quad ; \quad
\langle \dot{\sigma}_y \rangle =
       E_J \, \langle \sigma_z \rangle 
       - \sum_i^N v_i \, \langle b_i^{\dagger} b_i \sigma_x \rangle \quad ; \quad
\langle \dot{\sigma}_z \rangle 
      =  - E_J \, \langle \sigma_y \rangle 
\end{equation}
In the rhs averages of new operators which involve also the 
localized levels and the bands are generated and we can write new equations 
for them. This procedure could be iterated and we would obtain 
an infinite chain of equations, but instead  at this stage we factorize 
high order averages, so we are left with a set of $3(N+1)$ equations. 
In practice
we ignore the cumulants
$\langle b_i^{\dagger} b_i b_j^{\dagger} b_j \rangle_c$ and 
$\langle b_i^{\dagger} b_i b_j^{\dagger} b_j
\sigma _\alpha \rangle_c$ for $i\neq j$ and insert the relaxation dynamics  
for the BCs in the approximated terms. 
This method gives accurate results for general values of $g_i$ even if 
$v_i / E_J$ is not very small, as we checked by comparing with numerical 
evaluation of the reduced density matrix of the qubit  with one and two BCs.
Results are presented in Fig.\ref{figure:equil} 
where the time Fourier transform of  $\langle \sigma_z (t)\rangle$, 
proportional to the average charge on the island, is shown. 
We assumed factorized initial condition for the qubit and the BCs.
We first consider a set of weakly coupled BCs in the range 
$[\, 10^{-2}, 10 \,]\,  E_J$ 
which determine $1/f$ noise in a frequency interval around the operating 
frequency. 
The coupling strengths $v_i$ have been generated uniformly with approximately 
zero average and with magnitudes chosen in order to yield the 
amplitude of typical 
measured spectra~\cite{zorin96,nakamura-echo,kn:noise-set} 
(extrapolated at GHz frequencies).
These BCs are weakly coupled, and  
determine a dephasing rate which reproduces the prediction
of Eq.(\ref{eq:master-result}) 
(Fig.\ref{figure:equil} dotted line 
$\Gamma_{\phi}/E_J \approx 1.65 \times 10^{-4}$). 
Now we add a slower (and strongly coupled $g_0=v_0/\gamma_0=8.3$) BC, in order
to  extend the $1/f$ spectrum to lower frequencies.
The added BC gives negligible contribution to $S(E_J)$
so according to Eq.(\ref{eq:master-result}) it 
should not modify $\Gamma_{\phi}$. Instead, as shown in
Fig.\ref{figure:equil}a, we find that the strongly coupled BC alone 
determines a dephasing rate comparable with that of the weakly coupled
BCs. The overall $\Gamma_{\phi}$ is more than twice the prediction of 
Eq.(\ref{eq:master-result}). 

The above result shows that $\Gamma_{\phi}$ does not depend only on 
$S(E_J)$. Moreover we checked that information beyond the full 
$S(\omega)$ is needed: we considered sets of charges with different $N$ and 
$v_i$ which realize the same power spectrum $S(\omega)$ and we find
they yield different values of $\Gamma_{\phi}$. The decoherence is larger 
if BCs with $g \raisebox{-3pt}{\small $\,\gtrsim$} 1$ are present in the 
set. In summary these results show that Eq.(\ref{eq:master-result})
{\em underestimates} the effect of strongly coupled BCs and in particular that
a finite $\Gamma_{\phi}$ at the optimal point can be obtained even if 
the $1/f$ spectrum does not extend up to frequencies $\sim E_J$.

If we further slow down the added BC we find that $\Gamma_{\phi}$
increases toward values $\sim \gamma_0$, the switching rate of the BC. 
This indicates that the effect of 
strongly coupled BCs on decoherence tends to saturate. We discuss later 
similar results for the case of pure dephasing, where this conclusion 
can be made sharp, but we stress here that this is a consequence of the
nonlinearity inherent to the discrete nature of the BC environment.
In this regime we observe also 
effects related to the initial preparation of the strongly coupled BC (see 
Fig.\ref{figure:equil}b). As we will discuss later we may describe 
different experimental
procedures using different preparations of the environment. Thus we 
have here a first example of the fact that different measurement protocols 
should be analized separately as far as dephasing due to BCs is concerned.

\section{Pure dephasing} 
In  the absence of 
the tunneling term Eq.(\ref{eq:hamiltonian-backcharges}) is a model for pure 
dephasing. 
The analysis can be carried out using the exact result 
Eq.(\ref{eq:gamma-pure-dephasin-general}). We notice that for our model
Eq.(\ref{eq:hamiltonian-backcharges}) 
${\cal H}_\eta$ can be decomposed in a sum of commuting terms each referring
to a BC, so if we assume factorized $w_E(0)$  Eq.(\ref{eq:gamma-pure-dephasin-general}) also factorizes
$\rho_{01}(t) = \rho_{01}(0) 
\prod_{j=1}^{N} \exp\{-i (\varepsilon - v_j/2)t \} f_j(t)$ in averages 
referring to a single BC.
Using a real-time path-integral technique, the general form
of $f_j(t)$ in Laplace space is obtained~\cite{kn:PRL}
\begin{equation}
f_j(\lambda) =  \frac{
\lambda + K_{1,j}(\lambda) 
- i \, v_j / 2 \, \delta p_{j}^0} 
{\lambda^2 + \left ( v_j / 2 \right )^2 
+ \lambda K_{1,j}(\lambda) +
\, v_j / 2\,  K_{2,j}(\lambda)} \, ,
\label{eq:poles-exact}
\end{equation}  
where $\delta p_{j}^0 = 1 -2 \langle b_j^{\dagger}b_j \rangle_{t=0}$ specify
the initial conditions for the charges.
The kernels $K_{1,j}(\lambda)$ and $ K_{2,j}(\lambda)$
are expressed by formal series expressions in the tunneling amplitudes $T_j$.

An interesting explicit form of the kernels is obtained in the 
Non Interacting Blip Approximation (NIBA)~\cite{kn:weiss}, which amounts to
approximate the kernels $K_{1,j}(\lambda)$ and 
$K_{2,j}(\lambda)$ at lowest order in the tunneling amplitudes $T_j$.
The result is
$$
K_{1,j}(\lambda)= \gamma_j \quad ; \quad
K_{2,j}(\lambda) = - { \gamma_j \over \pi} 
[\psi (1/2 +  \beta/2 \pi(\lambda-i \epsilon_{cj})) 
- \psi (1/2 + \beta/2 \pi(\lambda + i \epsilon_{cj}))] \quad.$$ 
In order to appreciate the physical meaning of the NIBA result, we notice that
it is also obtained using the Heisenberg equations of motion, which yield
a closed set of equations provided that  the electronic band is assumed in 
thermal equilibrium. Thus the NIBA result, besides describing the dynamics 
of the qubit plus BC at all orders in the couplings $v_j$, is valid 
even if the couplings $T$ with the bands are not small. 
What is possibly missed are details of the 
dynamics of the bands, which is in total agreement with the phylosophy of 
the modified roadmap outlined in Sec.\ref{sec:model and methods}.

\subsection{Nearly incoherent BC dynamics}
\label{nearly}
We now want to apply these results to a set of BCs which produce $1/f$ noise. 
In order to produce a clear physical picture, we will consider only the 
physically relevant limit where the BCs have an incoherent dynamics. 
This limit can be formally obtained from the NIBA result by approximating 
the kernel
$K_{2,j}(\lambda) \sim K_{2,j}(0)$, and can be expressed
in analytic form
\begin{eqnarray} 
\label{offdiag}
\frac{ \rho_{01}(t)  }{ \rho_{01}(0)  }  &=&
\, e^{-i \varepsilon t} \,
\prod_{j=1}^{N}  \, e^{i v_j \, t/2} 
\left \{  A_j \, e^{-{\gamma_j \over 2} (1 - \alpha_j)t}
	\, + \,(1-A_j) \, e^{-{\gamma_j \over 2} (1 + \alpha_j)t} 
	\right \} 
\end{eqnarray}
where 
\begin{eqnarray}
\label{alpha} 
\alpha_j \,=\, \sqrt{1-  g_j^2
- 2i \, g_j	\, 
\tgh \left(\beta \varepsilon_{cj}/2 \right)} 
\qquad ; \qquad
A_j \,=\, {1 \over 2 \alpha_j} \, 
\left( 1+ \alpha_j - i \, \delta p_{j}^0 \, g_j \right) \, .
\end{eqnarray} 
In the framework of the NIBA this result is valid if
$\varepsilon_{ci} \, , v_i \, , \gamma_i \ll K_B T $, but again its validity
is broader. Indeed Eq.(\ref{offdiag}) it may be obtained
in different ways, for instance by analizing the system of qubit and BC by a 
master equation and taking the limit where $\varepsilon_{ci}$ is the 
largest scale, or even as the exact result of a model where 
the coupling operator 
$\sum_i v_i b^\dagger_{i} b_{i}$ is substituted  by a
classical stochastic 
process ${E}(t)$ which is the sum of 
random telegraph processes (see App.~\ref{sec:stochastic}).

The form of Eqs.(\ref{offdiag},\ref{alpha}) 
elucidates the different role of weakly
and strongly coupled BCs in the decoherence process. We focus on the
long time limit $\gamma t \gg 1$.
Dephasing due to each BC comes from the sum of two exponential terms.
For extremely weakly coupled charges, $g_j \ll 1$, whe have
$\alpha_j \approx 1$ and  only the first term is important, which
describes decay of the coherences with a rate
$\approx 1/[4 \cosh^2(\beta \varepsilon_{cj}/2)] \, v_j^2/\gamma_j$.
This is precisely the result of Eq.(\ref{eq:master-result}) for the 
adiabatic rate $\Gamma_\phi^0$.
For strongly coupled charges, $g_j \gg 1$, 
each of the two exponentials in Eq.(\ref{offdiag}) 
expresses roughly the same decay rate 
$\propto \gamma_j$, the switching rate of the individual BC, and moreover
they come with the same weight. The individual
$\alpha_j$s have large imaginary parts so 
the main effect of strong coupling with the qubit is not decay but rather
an energy shift.  The short time limit, which is more important in actual 
experiments, will be discussed in the next subsection.

The long time limit allows to draw a simple picture of the effect of a BC: 
the qubit
is practically insensitive to very fast 
BCs ($g_j \gg 1$), which are averaged completely. 
The dephasing effect increases
as $v_i^2/\gamma_i$ as the BC gets slower but eventually saturates. Indeed a 
very slow BC ($g_j \gg 1$) will dephase only when it switches (effect 
$\propto \gamma_i$) so for most of the time it provides a static extra 
polarization for the qubit. Based on this picture one could conclude that 
slow charges could be neglected, however it is not a priori clear if this is a
valid choice for the $1/f$ spectrum, where the number of slow charges is large
due to the $P(\gamma) \propto \/\gamma$ distribution. Eq.(\ref{offdiag}) allows
to answer to this question, but we defer this point to section 
\ref{sec:single-shot} and we first discuss in 
detail other aspects of the result for a single BC.

\subsection{Single BC}
\label{sec:single-BC}
\begin{figure}[htb]
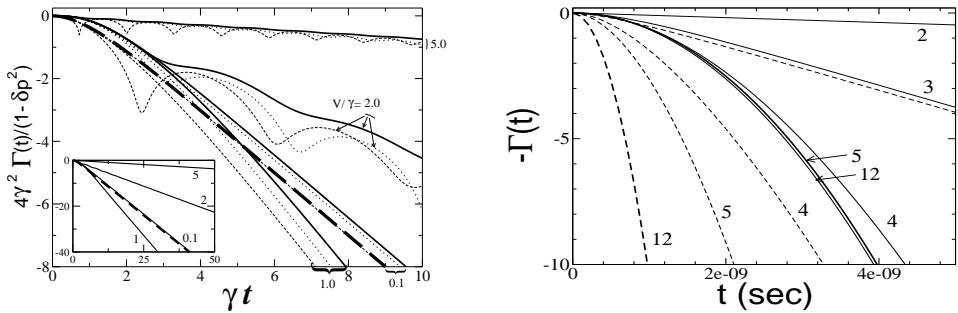

\vspace{0.3mm}
\begin{center}
\resizebox{!}{40.5mm}
{\includegraphics{decay-singlecharge-01-II.eps}}
\hspace{5mm}
\resizebox{60mm}{40.5mm}{\includegraphics{single-shot-1overf.eps}}
\end{center}
\caption{(a) Reduced $\Gamma(t)$ due to a BC with different preparations 
(thin lines) for the indicated values of $g$. Inset: longer time behavour 
for stable state preparation. The thick dashed line represents the oscillator 
approximation;
(b) Saturation effect of slow BCs for a 1/f spectrum.
Relevant parameters ($\overline{v} = 9.2 \times 10^7 Hz$, $n_d = 1000$)
give typical experimental measured noise levels and 
reproduce the observed decay of the echo signal~\protect\cite{nakamura-echo}
in charge Josephson qubits.
Couplings $v_i$ are distributed with  dispersion
$ \Delta v / \overline{|v|} = 0.2$. 
$\Gamma(t)$ is almost unaffected by strongly coupled charges 
(the label is the number of decades included). }
\label{fig5}
\end{figure}

Results for a single BC are concentrated in Fig.~\ref{fig5}a. We consider 
Eq.(\ref{offdiag}) for $N=1$ and 
look at the
quantity 
$$
\Gamma(t) = - \ln \left|
\frac{ \rho_{01}(t)  }{ \rho_{01}(0)  } 
\right|
$$ 
In Fig.\ref{fig5} we compare this quantity 
with the exact result for an oscillator environment,
Eq.(\ref{eq:gamma-pure-dephasing-oscillators}), 
which is reported as 
the thick dashed line. We show the effect of a single BC for various initial 
conditions and different values of $g$. Weakly coupled charges 
($g=0.1$) give a contribution close to $\Gamma_{osc}(t)$, whereas 
deviations are observed in the other cases. In particular BCs such 
that $g >1$ show slower dephasing compared to an oscillator environment with 
the same $S(\omega)$, as a manifestation of saturation effects. Recurrences 
at times comparable with $1/v$ are visible in $\Gamma(t)$. 

Fig.~\ref{fig5}a allows to discuss initial conditions of the BC. For each 
value of $g$ three cases are shown. The thick lines correspons to 
$\delta p^0 = 1$ (the dotted lines to $\delta p^0 = -1$), i.e. the 
BC is initially in the lower (higher) energy classical configuration. 
These are typical initial situations to be considered in a single shot 
process. In this case $\Gamma(t)$ describes dephasing {\em during} 
time evolution.
Differences in the behavior of $\Gamma(t)$ reflect memory effects. 
In particular for $\gamma t \ll 1$ the approximate behavior is 
$\Gamma(t) \approx  v^2 t^2 (1-\delta p_0^2)/8
- \gamma v^2 t^3 \, (1+2 \delta p_0 \overline{\delta p}-3 \delta p_0^2 )/24$,
whereas $\Gamma_{osc}(t) \approx v^2 t^2 (1- \overline{\delta p}^2)/8$.
Thus if  $\delta p^0 = \pm 1$, $\Gamma(t) \propto t^3$
which accounts for the fact that a two level system is stiffer than a set 
of oscillators.
On the other hand if we choose $\delta p^0 = \overline{\delta p}$, the 
equilibrium value, $\Gamma(t)$ decays more rapidly, 
$\Gamma(t) \approx \Gamma_{osc}(t)$. 
These latter choice of initial conditions corresponds physically
to repeated measurements in which we do not control the preparation of the BC
so $\Gamma(t)$ describes both dephasing {\em during} 
time evolution and the blurring of the total signal of several realizations
of the time evolution with slightly different characteristic frequency, 
a sort of inhomogeneuos broadening. Notice finally that if we use the above
equilibrium
initial conditions  $\Gamma(t)$ 
follows $\Gamma_{osc}(t)$ for short times, the two curves becoming
indistinguishable only if moreover $g \ll 1$. This observation is important
for the subsequent analysis of inhomogeneuos broadening (Sec.~\ref{sec:inhomogeneuos}).

The above analysis of decoherence due to a single BC has clearly
evidenced the different qualitative influence on the qubit dynamics
of weakly and strongly coupled BCs.
A single weakly coupled, $v/\gamma \ll 1$, BC behaves as a source of 
gaussian noise. Thus decoherence only depends on the power spectrum
of the fluctuator, and does not show any dependence on the initial
condition of the BC.
On the other side decoherence due a single strongly coupled BC, 
$v/\gamma \gg 1$, displays saturation effects and dependence on the
initial condition.

\section{$1/f$ noise in single shot measurements}
\label{sec:single-shot}
We now want to check how the intuitive picture for an environment made of
a single fluctuator may apply to a set of charges producing $1/f$ noise.
The distribution of BCs involves both weakly
and strongly coupled fluctuators, so that no typical time scale is
present.
Moreover since a  very large number of slow fluctuators is present, it is not
a priori clear how saturation effects will manifest.

In Fig.\ref{fig5}b we 
show the results for a sample with a number of BCs per decade
$n_d=1000$ and with $v_i$ 
distributed with small dispersion around the value 
$\overline{|v|} = 9.2 \times 10^7 Hz$. We choose initial conditions 
$\delta p_j^0 = \pm 1$ randomly distributed on the set of $N$ charges to 
reproduce equilibrium conditions in the ensemble, 
$(1/N) \sum_j \delta p_j^0 \approx \overline{\delta p}$. We checked 
that
for every realization of the initial conditions we obtained roughly the
same overall polarization and the same dephasing. What we are going to 
calculate is then the result of single shot measurements, i.e. the average 
signal of several experiments on the time evolution of the qubit where 
the total initial polarization of the environment is recalibrated 
before each experiment. This quantity is essentialy dephasing {\em during} 
time evolution and the corresponding protocol is such to minimize the effects 
of the environment. To make reference to a concrete situation, the results we
present were calculated with parameters close to the 
experiments~\cite{nakamura-echo}.

In order to illustrate the different role played by the BCs with 
$g_j \ll 1$ and $g_j \gg 1$, 
we now perform a spectral analysis of the effects of the environment.
We consider  sets of BCs with the same 
$\gamma_M = 10^{12} Hz$\footnote{This value of $\gamma_M$ is surely too 
large, but in this section we are interested to the way dephasing changes by 
adding low-frequency noise decade after decade} 
and decreasing $\gamma_m$, all producing $1/f$ noise
with the same amplitude $\cal A$ in the corresponding frequency range. 
Solid lines in Fig.\ref{fig5}b represent $\Gamma(t)$ calculated
with Eq.(\ref{offdiag}). 
In this example the dephasing is given by BCs 
with $\gamma_j > 10^7 Hz \approx \overline{|v|} /10$. 
The main contribution comes from three decades at frequencies around 
$\overline{|v|}$. The overall effect of the strongly coupled BCs 
($\gamma_j < \overline{|v|} /10$) is minimal, despite of their large 
number, showing the saturation effect of low-frequency noise. 
Dashed lines represent $\Gamma_{osc}(t)$, the result
of the approximation of the environment with a set of harmonic 
oscillators, Eq.(\ref{eq:gamma-pure-dephasing-oscillators}). 
In this case low-frequency noise does not saturate,
each decade of noise equally contributes to dephasing and  $\Gamma_{osc}(t)$
depends on the low-frequency cut-off of $1/f$ noise. We notice finally that
for this single shot protocol $\Gamma(t)$ is roughly given by $\Gamma_{osc}(t)$
provided we use $\omega \sim \overline{|v|}$ as a low frequency cut-off.

This observation also explains the fact that our results are not very 
sensitive to the value of $n_d$ we choose. Indeed in order to reproduce 
a given amplitude $\cal A$ we must keep constant for each decade 
$\sum_i v_i^2 \approx n_d \overline{v^2}$, meaning that the 
effective low-frequency cut-off $\sim \overline{|v|}$ varies 
as $n_d^{-1/2}$. For $n_d \to \infty$ the low-frequency cut-off goes to
zero, as in the result Eq.(\ref{eq:gamma-pure-dephasing-oscillators}) 
for the oscillator environment. 

Finally we point out that even if we performed our calculation by assigning
all the microscopic parameters of the environment, a reliable estimate of 
dephasing for single shot measurements turns out to depend on a single 
additional parameter besides the power spectrum $S(\omega)$, namely 
the average coupling  $\overline{|v|}$ or equivalently the number 
of charges producing a decade of noise, $n_d$.

\section{Comparison with the oscillator environment}
\label{sec:comparison}
In this section we make a more quantitative comparison of our results with 
the results for an equivalent oscillator environment.
It is useful to consider our result from the point ov view of the classical
stochastic approach of App.~\ref{sec:stochastic} which expresses the off 
diagonal element of the reduced density matrix of the qubit as an average over 
{\em classical} stochastic processes generated by a set of random telegraph 
fluctuators. The result for {\em quantum} oscillators 
Eq.(\ref{eq:gamma-pure-dephasing-oscillators}) is the second cumulant 
expansion of  Eq.(\ref{offdiag}). Alternatively one may verify that
$$
\Gamma_{osc}(t) = {1 \over 2} \, \sum_i g_i^2 
\left[ \partial^2 \Gamma(t) \over \partial g_i^2 \right]_{g_i=0}
$$
where $\delta p_i^0 = \overline{\delta p}$ has been posed in $\Gamma(t)$.
This result is intuitive in that in the limit $v_i \to 0$ the noise is 
produced by $n_d \to \infty$ fluctuators per decade, and its discrete 
nature is lost. This agrees with the fact that in this limit all BCs are
weakly coupled.

To give an idea of the numbers involved we checked this conclusion by
calculating $\Gamma(t)$ from Eq.(\ref{offdiag}), using 
different sets of BCs. The results are shown in
Fig.\ref{gaussian}a. 
The power spectrum $S(\omega)$ is identical for 
all the curves shown, which differ in the choice of  $n_d$.
The gaussian behavior is recovered in the long time limit 
$\gamma_m t \gg 1$ for large enough $n_d$ (all the BCs are weakly coupled).
If in addition we take $\delta p_j^0 = \overline{\delta p}$,  
$\Gamma(t)$ approaches $\Gamma_{2} (t)$ also at short
times.

\section{Repeated measurements and inhomogeneuos broadening}
\label{sec:inhomogeneuos}
Measurements of the state of a charge-Josephson qubit involve the measurement
of a single extra Cooper pair in the superconducting island. Different 
strategies of measurements have been proposed and experimental prototypes
have been produced, but in practice measurements are not single-shot.
In the simple scheme used in Ref.~\cite{kn:Nakamura1,nakamura-echo} the time 
evolution  procedure
is repeated $\sim 10^{5}$ times and what is measured is 
the total current due to the possible presence of the extra Cooper pair 
in all the repetitions. The signal is then the sum over different possible
time evolutions of the BCs with initial conditions which are also randomly 
fluctuating. The average effective precession frequency of the qubit is 
then different in different realizations and the overall signal decays faster 
than a single shot measurement would yield. In this respect this additional
decay of the signal is analogous to inhomogeneuos broadening in NMR, where
the signal is collected by many noninteracting ``qubits'' each with its 
specific environment
which determines different average precession frequencies. One difference 
may possibly be that strong correlations may exist between the different
repetitions, due to the strong correlations in time of $1/f$ noise.
\vspace{3.5mm}
\begin{figure}[bht]
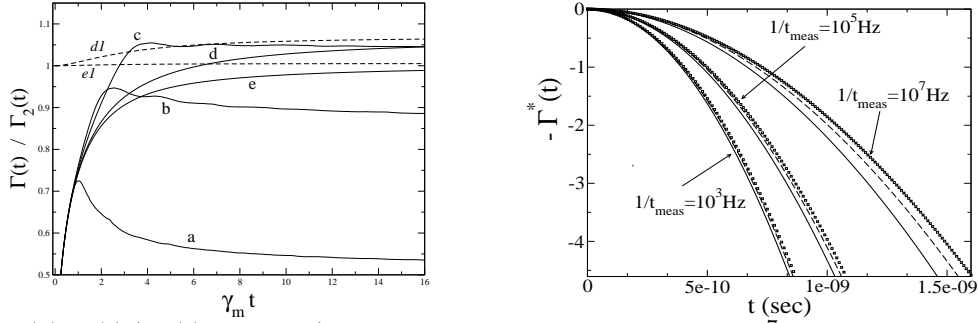

\centerline{
\resizebox{!}{42mm}
{\includegraphics{fig4.eps}}
\hspace{10mm}
\resizebox{60mm}{42mm}
{\includegraphics{fid.eps}}}
\caption{
(a) $\Gamma(t)/ \Gamma_2(t)$ for a $1/f$ spectrum between 
$\gamma_{m}=2 \times 10^7$ and $\gamma_{M}=2 \times 10^9$ with
different numbers of BCs per decade:
(a) $n_d = 10^3$, (b)  $n_d = 4 \times 10^3$,\
(c) $n_d = 8 \times 10^3$, (d) and ({\it d1}) $n_d = 4 \times 10^4$,
(e) and ({\it e1}) $n_d = 4 \times 10^5$.
Full lines corresponds to $\delta p_j^0= \pm 1$, dashed lines to 
equilibrium initial conditions for the BCs.
(b) Different averages over $\delta p^0_j$
for 1/f spectrum reproducing the measured noise level of 
Ref.~[20]:
 $\overline{|v|} = 9.2 \times 10^6 Hz$, 
$n_d=10^5$, $\gamma_m = 1 Hz$, $\gamma_M = 10^9 Hz$. 
Dashed lines correspond to the oscillator approximation with 
a lower cut-off at $\omega = \min \{\overline{|v|} , 1/t_m \}$.}
\label{gaussian}
\end{figure}
In order to study the overall signal we have to sum Eq.(\ref{offdiag}) over 
different
sets of $\{\delta p_{j\alpha}^0\}$ representing initial conditions for each
repetition starting at $t=t_\alpha$. Equivalently we may average 
Eq.(\ref{offdiag})
over a suitable distribution of $\{ \delta p_j^0 \}$ which depends on time.
If BCs are assumed to be independent this distribution factorizes and since
$\{\delta p_j^0\}$ enters linearly the coefficient $A$ we have only to 
evaluate Eq.(\ref{offdiag}) with $\delta p_j^0 = \delta p_j^0(t_m)$, where 
$\delta p_j^0(t_m)$ is the average of the values of  $\delta p_j$ sampled
at regular times $t_\alpha$ for the overall time $t_m$. As a rough estimate 
we may let $\delta p_j^0(t_m)=\delta p_j^0(0)= \pm 1$ 
if $\gamma_j t_m < 1$ and 
$\delta p_j^0(t_m)= \overline{\delta p}$ for $\gamma_j t_m > 1$, i.e. slow
charges do not change roughly initial condition (but still they may dephase
during time evolution) whereas fast charges completely average during $t_m$.
In this way the additional scale $t_m$ enters the problem. 

We consider only the case of long overall measurement time 
$\overline{|v|} t_m \gg 1$ which is 
pertinent to the present day experimental conditions. 
From the results of Sec.~\ref{sec:single-BC} and Sec.~\ref{sec:comparison}
we may infer that BCs with $\gamma < 1/t_m < \overline{|v|}$, being 
strongly coupled and with $\delta p_j^0= \pm 1$, are ineffective  
whereas for the other BCs, being averaged, we may take 
$\Gamma^{(i)}(t) \approx \Gamma^{(i)}_{osc}(t)$ for small enough times
and, as a result, 
$\Gamma(t) \approx \int_{1/t_m}^{\infty} \hskip-2pt d\omega \;
S(\omega)(1 - \cos \omega t)/(\pi \omega^2)$. 
This would proof the recipe proposed in Ref.~\cite{kn:cottet}. 
It is intersting to notice that this 
result and possible extensions to higher order effects~\cite{kn:shnirman} in 
the qubit-environment can be obtained with no reference to the quantum 
nature of the environment, and depend on the classical statistical properties
of an equivalent random process. 

In Fig.\ref{gaussian}b, (dotted lines) we show that indeed 
dephasing calculated as outlined
above, is roughly given at short times by the oscillator environment 
approximation with a lower cut-off taken at 
$\omega \approx  \min \{\overline{|v|}, 1/t_m\}$ (dashed line). 
We also show results with a different averaging procedure 
$\delta p_j^0(t_m)= 1/t_m \int_0^{t_m} dt 
\overline{\delta p(t)}$
which takes 
into account the strongly correlated dynamics of $1/f$ noise 
(solid lines in Fig.\ref{gaussian}b,). 
These correlation do not affect the results except possibly for 
$t_m \approx \overline{|v|}$.

\begin{figure}[bht]
\centerline{
\resizebox{69mm}{40mm}{\includegraphics{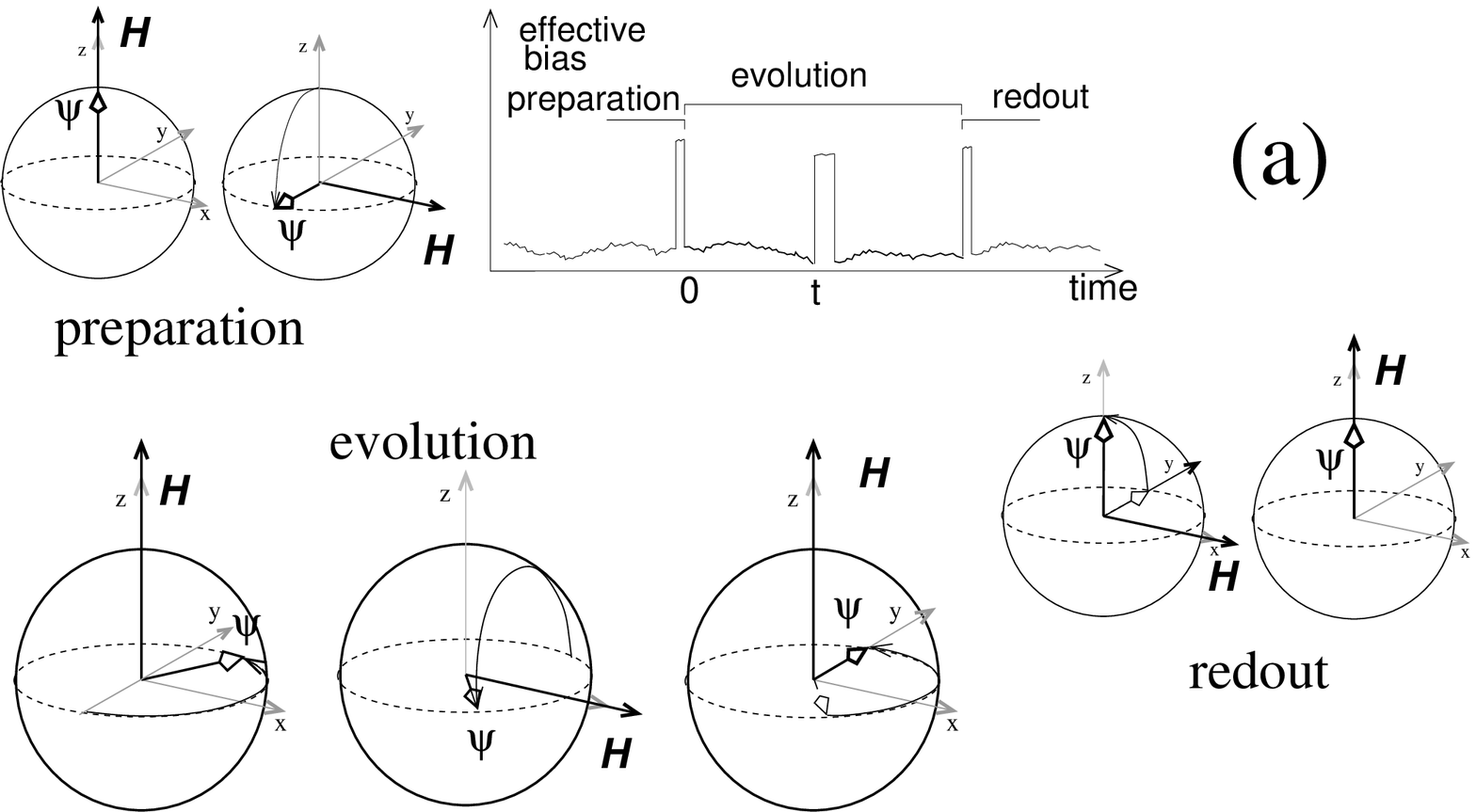}}
\hspace{2mm}
\resizebox{!}{40mm}{\includegraphics{ecoLT.eps}}
}
\caption{(a) Bloch sphere descritpion of the echo protocol.
(b) Decay of the echo amplitude for a $1/f$ spectrum, with noise level
as in Fig.(\ref{fig5}b) with $\gamma_m = 1Hz$ and different $\gamma_M$. 
The dashed lines correspond to the oscillator environment approximation.}
\label{fig:echo-shot}
\end{figure}

\section{Charge echo}
Echo-type techniques have been suggested~\cite{kn:Nakamura1,kn:cottet} and 
experimentally tested~\cite{nakamura-echo} as a tool to reduce 
inhomogeneous broadening due to the low-frequency fluctuators of the 
$1/f$ spectrum. In the experiment of Ref.~\cite{nakamura-echo} the echo 
protocol consists of a $\pi/2$ preparation pulse, 
a $\pi$  swap pulse and a $\pi/2$ measurement pulse. Each pulse is separated
by the delay time $t$ (see Fig.~\ref{fig:echo-shot}). Since the duration of
each pulse is negligible with respect to $t$ we can estimate the 
decay of the coherence only looking at the evolution during the delay times. 
Within the semiclassical approach we obtain~\cite{kn:mqc2} 
\begin{eqnarray}
\label{eq:decoherence2}
{\rho_{01}(t) \over \rho_{01}(0)} &=& \prod_j \Bigl\{
A^1_j  \mathrm{e}^{-(1-Re \alpha_j)\gamma t} + A^2_j  
\mathrm{e}^{-(1+Re \alpha_j)\gamma t}\\ && \hskip50pt
+ A^3_j  \mathrm{e}^{-(1- i Im \alpha_j)\gamma t} + A^4_j 
\mathrm{e}^{-(1+i Im \alpha_j)\gamma t}
\Bigr\}
\nonumber
\end{eqnarray}
where $\alpha_j$ is given by Eq.(\ref{alpha}) and 
information on $\delta p^0_j$ is contained in $A_j^k$. 
We are interested to 
$\Gamma^{(2)}(t) = - \ln \left| {\rho_{01}(t) / \rho_{01}(0)}
\right|$.
Again the expansion of Eq.(\ref{eq:decoherence2}) to the second cumulant
gives the result for an equivalent environment of quantum oscillators
$$
\Gamma^{(2)}_{osc}(t) = {2 \over \pi}
\int_{0}^{\infty} \hskip-2pt d\omega  \;
S(\omega) \;
{(1 - \cos \omega t)^2 \over \omega^2} 
$$

Results are shown
in Fig.~\ref{fig:echo-shot} for the same set of BCs of 
Fig.~\ref{fig5}
whose parameters are 
close\footnote{As it is clear from the results of Sec.~\ref{sec:single-shot},
for a given measured $\cal A$  
we had to infer $n_d$. This we made by comparing our results of this
section with the echo experiment. This allows only a rough estimate of 
$n_d$ because, as in the other cases, the dependence on $n_d$ is weak.}
to the experimental situation of Ref.~\cite{nakamura-echo}.
First of all we checked that the dependence of $\Gamma^{(2)}(t)$ on 
initial conditions $\delta p^0_i$ is extremely weak even if we 
start from out of equilibrium configurations of the set of BCs.
Moreover  $\Gamma^{(2)}(t) \approx \Gamma^{(2)}_{osc}(t)$. This 
means that the echo procedure actually cancels the effect of 
strongly coupled charges\footnote{This conclusion is valid as long as 
the delay time is short, $t \overline{|v|} \ll 1$}. Only weakly 
coupled charges contribute to $\Gamma^{(2)}(t)$ and the environment
behaves as a suitable set of harmonic oscillators.
We remark that in the regime of parameters we
consider, for given noise amplitude the echo signal is strongly dependent 
on the high frequency cut-off, as studied in  Fig.~\ref{fig:echo-shot}.

\section{Conclusions}

In conclusion we have studied dephasing due to charge fluctuations in 
solid state qubits. We analized and compared two models for the environment,
namely an environment of harmonic oscillators and an environment made of
fluctuators with applications to $1/f$ noise which is probably the most 
serious limitation for these devices. Even if decoherence comes in general 
from the entanglement of the system with its environment, 
for solid state devices 
measurements of system-environment correlations are extremely hard and
the reduction of the amplitude of the coherent signal in specific experiments 
may often be studied in less fundamental terms. We discussed the problem of
the information needed to characterize the effect of the environment. 
For environments which are weakly coupled and fast or environments 
made of harmonic oscillators the information needed is entirely 
contained in the power spectrum of the operator which couples the environment 
to the system. For a fluctuator environment with $1/f$ spectrum memory 
effects and higher order moments are important so additional information
is needed. We discussed in 
detail the remarkable case of pure dephasing where the reduction of the 
amplitude of the coherent signal is given by correlation functions of the 
environment alone (see Eq.(\ref{eq:gamma-pure-dephasin-general})) and provided
exact results for the fluctuator environment. In this case the additional 
information of the environment needed depends on the protocol but often
reduces to a single parameter. A new energy scale emerges, the average coupling
$\overline{|v|}$ of the qubit with the BCs, which is the additional 
information needed to discuss single shot experiments (alternatively one 
should know the order of magnitude of $n_d$, the number of BCs per decade 
of noise). For repeated experiments the relevant scale is instead 
$ \min \{\overline{|v|}, 1/t_m\}$ where $t_m$ is the overall
measurement time. Finally echo measurements are sensitive to the high-frequency
cut-off $\gamma_M$ of the $1/f$ spectrum.

Our results are directly applicable to other implementations of solid state 
qubits. We only mention Josephson flux qubits~\cite{kn:tian00,kn:mooji} wich 
suffer from 
similar $1/f$ noise, originated from trapped vortices. In that case the 
eigenstates of $\sigma_z$ are the flux states of the device.
Also the the parametric effect of $1/f$ noise on the coupling energy of a 
Josephson junction~\cite{kn:clark} can be analized within our model, as long as
individual fluctuators do not determine large variations of $E_J$. In this
case it may be possible that the same sources generate both charge noise 
and  fluctuations of $E_J$. This can be accounted for in our model by 
choosing the ``noise axis'' as the $\hat{z}$ axis. 

In this work we have also briefly discussed the possibility of finding
optimal operating points for the qubit. This idea has been succesfully 
implemented in the experiment of Ref.~\cite{kn:vion} and consists in operating
with external parameters tuned at the points where the energy splittings of
the system are less sensitive to fluctuations. However a slight deviation 
from the optimal point determines a strong degradation 
of the performances. An equivalent point of view 
is that the system should be tuned at the point where adiabatic dephasing 
cancels in lowest order, since the effect of a low frequency environment
(in particular the $1/f$ environment) is minimized. An explicit example 
is the recent proposal of implementing a communication protocol with 
Josephson junctions~\cite{kn:plastina}. 
Low-frequency noise can also be minimized by echo techniques, but the 
flexibility in the implementation of gates is greatly reduced. A possibility
is to implement quantum computation using Berry phases~\cite{kn:nature}, 
where the design of gates includes an echo procedure, but a detailed 
analysis of the effect of $1/f$ noise is still missing. 

Finally we mention that the sensitivity of coherent devices may be used 
to investigate high frequency noise~\cite{kn:kouvenowen}. In particular 
an accurate matching between measured inhomogeneous broadening,  
echo signal and relaxation may give reliable information on the actual 
existence of BCs at $GHz$ and on the high-frequency cut-off of the $1/f$ 
spectrum.

\appendix
\section{Linear quantum noise} 
Fluctuations of the electromagnetic circuit can be modeled by coupling the 
system to an environment of harmonic oscillators~\cite{kn:weiss,kn:dissipative}
which mimicks the external impedences (see Fig.\ref{fig:cjqubit}). 
In this Appendix we present a model for the electromagnetic environment and we 
derive an effective Hamiltonian $H_{eff}$ using no phenomenological 
argument~\cite{kn:LTcircuit}. 
This has two motivations. First, since in principle bare circuit parameters 
are well defined and tunable, we want to know precisely how this reflects 
on $H$.
Second, in general coupling to the environment produces decoherence and energy
shifts, which may in principle be large. In dissipative quantum 
mechanics shifts are usually treated either by introducing 
counterterms\cite{kn:dissipative} or by writing $H_{eff}$ 
in terms of renormalized quantities\cite{kn:dissipative}.
The role of induced shifts, which is minor in 
the devices of Refs.\cite{kn:Nakamura1,kn:vion}, may be crucial 
in various situations (e.g. geometric quantum computation\cite{kn:nature}, 
dynamics of registers and error correction devices).

We consider the Cooper pair box~\cite{kn:rmp} of Fig.~\ref{fig:cjqubit}. 
The external impedance is modeled by a suitable $LC$ transmission line
and the Lagrangian of the system is 
$${L} \;=\; \sum_{i=1,2} {C_i \dot{\phi}_i^2 \over 2} 
	- V_J(\phi_1) 
+ 
	\sum_\alpha \bigl[ 
	{C_\alpha \dot{\phi}_\alpha^2 \over 2} - 
	{\phi_\alpha^2 \over 2 L_\alpha}
	\bigr] 
$$
where $\dot{\phi}$ are voltage drops and the Josephson energy is
$V_J(\phi_1) =	- E_J  \cos (2 e \phi_1/ \hbar) $.
The environment is fully specified by the elements $C_\alpha$ and 
$L_\alpha$. The circuit is introduced by the constraint
$\dot{\phi}_1 + \dot{\phi}_2 + 	\sum_\alpha 
\dot{\phi}_\alpha = V_x$, which allows to eliminate one variable and
to write

\begin{eqnarray}
{L} &=& {C_\Sigma\, \dot{\eta}^2 \over 2} - 
	V_J\bigl(2 e (\kappa_2 \Phi -  \eta) \hbar \bigr)
	  + L_b
\nonumber\\
 L_b &=&{1 \over 2}\, C_e\, \dot{\Phi}^2 \,+ \,
	\sum_\alpha \bigl[ 
	{C_\alpha \over 2} \; \dot{\phi}_\alpha^2 \,- \, 
	{1 \over 2 L_\alpha} \; \phi_\alpha^2
	\bigr]
\nonumber
\end{eqnarray}
where $\eta =  \kappa_2 \phi_2 -  \kappa_1 \phi_1$, 
$C_{\Sigma}= \sum_{i} C_{i}$, $C_e=C_{1}C_{2}/C_{\Sigma}$, 
$\kappa_{1,2} = C_{1,2}/C_{\Sigma}$, and 
$\dot{\Phi} = V_x - 	\sum_\alpha  \dot{\phi}_\alpha$. We next 
diagonalize $L_b$ and obtain the form
$$
{ L}_b  \;=\; 
\sum_\alpha \Bigl[ 
	{m_\alpha \over 2} \; \dot{x}_\alpha^2 \,- \, 
	{m_\alpha \omega^2_\alpha \over 2} \; x_\alpha^2
	\Bigr]\,- \, C_e V_x \, \sum_\alpha  d_\alpha \dot{x}_\alpha
$$
In order to determine the parameters, notice that
$L_b$ describes a series $C_e-Z$ circuit, with 
$\sum_\beta  d_\beta \dot{x}_\beta = \sum_\alpha  \dot{\phi}_\alpha = V_Z$. 
By comparing the linear response with the known classical dynamics of $V_Z$ 
we determine the spectral density ($\omega > 0$)
$$
J^\prime(\omega)
= \sum_{\alpha} \;  { \pi d^2_{\alpha} \delta(\omega - \omega_\alpha)
\over 2 m_\alpha \omega_\alpha}
=
\mathrm{Re} \Bigl[ {Z(\omega) / \omega  \over 1 + i \omega Z(\omega) C_e}
\Bigr]
$$
Notice that since $L_b$ is quadratic this procedure is an exact way to 
perform the diagonalization, due to circuit theory, and is valid after 
quantization due to the Ehrenfest theorem. 
We then perform a (canonical) transformation 
$\chi = \eta - \kappa_2 \Phi$ and obtain the total Lagrangian
$L= L_a + L_b$ with
$$
{L}_a = {C_\Sigma \over 2}
	\bigl( \dot{\chi} + \kappa_2  V_x - 
	\kappa_2 \sum_\alpha  d_\alpha \dot{x}_\alpha
	\bigr)^2 
	+ 
	V_J( 2 e \chi/ \hbar)
$$
One can verify that the variable canonically conjugated to $\chi$ is the 
charge $Q$ in the island. The Hamiltonian corresponding to $L$ reads

\begin{eqnarray}
{H} 
&=&
{Q^2 \over 2 C_1} \; \,+\, Q \; \kappa_2
\sum_\alpha   { d_\alpha \over m_\alpha}  \; p_\alpha 
\,-\, E_J \, \cos \Bigl( {2 e \over \hbar} \chi \Bigr) 
\nonumber\\ && \hskip10pt  +   
\sum_\alpha \Bigl[ {p_\alpha^2 \over 2 m_\alpha} \,+\, 
 {m_\alpha \omega^2_\alpha \over 2}   \, x_\alpha^2 \Bigr] 
 \,+\,  C_e V_x \; \sum_\alpha {d_\alpha  \over m_\alpha} \; p_\alpha
\nonumber
\end{eqnarray}
where $p_\alpha$ are conjugated to $x_\alpha$ and we used the relation
$\sum_\alpha d_\alpha^2 / m_\alpha = 1/C_e$. A further canonical 
transformation of the environment 
($
\tilde{x}_\alpha = {p_\alpha/( m_\alpha \omega_\alpha)}$,  
$\tilde{p}_\alpha = - m_\alpha \omega_\alpha  x_\alpha$) yields

\begin{eqnarray}
\label{eq:ZC-superc-box-hamiltonian-3}
{H}_{eff} &=& {Q^2 \over 2 C_1} + V_J\bigl( {2 e \over \hbar} \chi \bigr) +
 \kappa_2 Q \sum_\alpha c_\alpha \, \tilde{x}_\alpha 
\nonumber\\ &+&
\sum_\alpha \Bigl[ 
	{\tilde{p}_\alpha^2 \over 2 m_\alpha} \,+ \, 
	{m_\alpha \omega^2_\alpha \over 2} \; \tilde{x}_\alpha^2
	\Bigr]\,+ \, C_e V_x \, 
\sum_\alpha c_\alpha \, \tilde{x}_\alpha
\nonumber
\end{eqnarray}
where
$c_\alpha = {d_\alpha \omega_\alpha}$ and we introduce 
$J(\omega)
= \pi  \sum_{\alpha} \delta(\omega - \omega_\alpha) c^2_{\alpha}/ 
(2 m_\alpha \omega_\alpha) = \omega^2 \,J^\prime(\omega)$, 
the spectral density. 

If we isolate the dc bias  $V_x(t) =V_x+ \delta V_x(t)$
and redefine 
$\tilde{x}_\alpha + c_\alpha C_e V_x/(m_\alpha \omega^2_\alpha) 
\, \to \, x_\alpha$
we finally obtain

\begin{eqnarray}
\label{eq:ZC-superc-box-hamiltonian-4}
{H}_{eff} &=& {Q^2 \over 2 C_1} - \kappa_2 V_x  Q + 
V_J\bigl( {2e \over \hbar} \chi \bigr)
+ Q  \kappa_2 \sum_\alpha c_\alpha  x_\alpha 
\nonumber\\
&+&
\sum_\alpha \Bigl[ 
	{p_\alpha^2 \over 2 m_\alpha} +  
	{m_\alpha \omega^2_\alpha \over 2} x_\alpha^2
	\Bigr] +  C_e \delta V_x(t)  
\sum_\alpha c_\alpha  x_\alpha \quad
\nonumber
\end{eqnarray}
Notice that the capacitance $C_1$ (and not $C_\Sigma$) enters the $Q^2$ term. 
and if we put $c_\alpha=0$ we do not obtain the Cooper pair box 
hamiltonian. This is correct because the 
environment represents global fluctuations of the circuit, not only of $Z$.
Notice that a static $Q$ shifts
the equilibrium points of the oscillators and also produces a $Q$-dependent 
shift of the zero of their energies. This latter can be reabsorbed 
in the charging energy if we write the oscillator hamiltonian using the 
shifted values $x_\alpha + c_\alpha \kappa_2  Q / (m_\alpha \omega^2)$, which
produces the extra term 
$- \kappa_2^2  Q^2 \sum_\alpha {c_\alpha^2 / (2  m_\alpha \omega_\alpha^2)}
= - Q^2/(2C_e)$ and 

\vspace{-12pt}
\begin{eqnarray}
\label{eq:ZC-superc-box-hamiltonian-5}
{ H}_{eff} &=& {Q^2 \over 2 C_\Sigma} - \kappa_2 V_x  Q + 
V_J\bigl( {2e \chi \over \hbar}\bigr) 	+  C_e \delta V_x
\sum_\alpha c_\alpha  x_\alpha
\nonumber\\&+&
\sum_\alpha \Bigl[ 
	{p_\alpha^2 \over 2 m_\alpha} +  
	{m_\alpha \omega^2_\alpha \over 2}  
	\Bigl(x_\alpha + {c_\alpha \kappa_2  Q \over  m_\alpha \omega^2}  
	\Bigr)^2
	\Bigr]
\end{eqnarray}
which reduces to the non dissipative form by letting $c_\alpha=0$.
This is a convenient starting point for a weak coupling analysis also 
because a static shift in the equilibrium points 
of the  oscillators has no effect even if it
is large.

\section{Dephasing due to classical stochastic fluctuations}
\label{sec:stochastic}

In the semiclassical approach 
the coupling operator 
$\sum_i v_i b^\dagger_{i} b_{i}$ is substituted  by a
classical stochastic 
process ${E}(t)$ which is the sum of random telegraph processes. 
The system is initially in a given superposition of eigenstates of ${\cal H}_Q$
Eq.(\ref{eq:qubit-hamiltonian}), say  
$\ket{\psi, 0} = \alpha \, \ket{a} + \, \beta \, \ket {b}$. We make now a first
assumption, namely we neglect 
relaxation and we write the time evolution using the adiabatic
theorem
$\ket{\psi, t} = \alpha \, \ket{a, t} + \, \beta \, 
\mathrm{e}^{-i \int_0^t dt^\prime \Omega(t^\prime)} \,\ket {b, t}$, where now 
we use the istantaneuos basis of 
${\cal H}_{ad} = {\cal H}_Q - {1 \over 2} E(t) \sigma_z$ and the 
instantaneuos level splitting 
$\Omega(t) = \sqrt{[\varepsilon + E(t)]^2+ E_J^2}$. The error introduced by
the random field $E(t)$ is mostly due to the phase factor and we make a second
assumption, namely we neglect
the difference between the istantaneuos basis and the eigenbasis of 
${\cal H}_Q$. In this case the diagonal elements of the density matrix
$\ket{\psi, t}  \bra{\psi, t}$ are constant whereas the off diagonal matrix
element is 
\begin{equation}
\label{eq:semi1}
\rho_{ba}(t)=\rho_{ba}(0) \, 
\mathrm{e}^{-i \int_0^t dt^\prime \Omega(t^\prime)}  \, .
\end{equation} 
At this stage the standard assumption
of the semiclassical approach is that averages over the environment 
correspond to averages over the stochastic process $E(t)$. Thus decoherence
during time evolution is estimated by
\begin{equation}
\label{eq:semi2}
\Gamma (t) = - \ln \left|
\overline{\mathrm{e}^{-i \int_0^t dt^\prime \Omega(t^\prime)}}  
\right| \, ,
\end{equation} 
where the overline denotes an average over different stochastic processes 
with given initial conditions. The effect of inhomogeneuos broadening is 
included by a further averages on initial conditions.

Let us discuss the validity of this result. If $E_J = 0$ the two 
assumption leading to Eq.(\ref{eq:semi1}) are rigorously verified and  the 
result is exact. By working out the average 
we shall pervent to the result Eq.(\ref{offdiag}). 
In the general case the first assumption is valid if $E(t)$
is slow on the time scale set by $\Omega$ and the second if $E(t) \ll \Omega$.
The condition of slow environment is equivalently expressed 
by requiring that $S(\omega)$ is small at frequencies $\omega \sim \Omega$. 
In this case relaxation in leading order (see Eq.(\ref{eq:master-result})) 
is negligible. Then Eq.(\ref{eq:semi2}) 
describes the effect of a large part of the degrees of freedom composing
the $1 / f$ environment, if not all of them.

We may apply Eq.(\ref{eq:semi2}) to repeated measurements as described in 
Sec.\ref{sec:inhomogeneuos}. If one assumes 
that for small times we can approximate the environment to a 
suitable set  of harmonic oscillators then we may approximate 
Eq.(\ref{eq:semi2}) to the second cumulant. Moreover, under the
assumption of small $E(t)$ it is easy to derive an expression for 
the decay of the off diagonal matrix element
$$
\Gamma (t) \;=\; 
\cos^2\theta \; \Gamma_{osc}(t) \,+\,  {\sin^4\theta \over 8 \Omega}
\int_0^t d t^\prime d t^{\prime \prime} \left[
\overline{E^2(t^\prime) E^2(t^{\prime \prime})} - \overline{E^2}^2
\right]
$$
This expression has been also derived in Ref.~\cite{kn:shnirman} where the 
oscillator environment has been studied in detail and we will not pursue this 
analysis. We only stress that the
correction to the term containing $\Gamma_{osc}(t)$ depends on the so
called second spectrum~\cite{kn:weissman} of the environment which is known 
to differ for gaussian and non gaussian stochastic processes. So if the 
correction is important, i.e. if the system turns out to be sensitive to the 
second spectrum, then the approximation of Eq.(\ref{eq:semi2}) to the second 
cumulant is inaccurate. 

\begin{figure}
\begin{center}
\raisebox{15mm}{\resizebox{!}{20mm}{\includegraphics{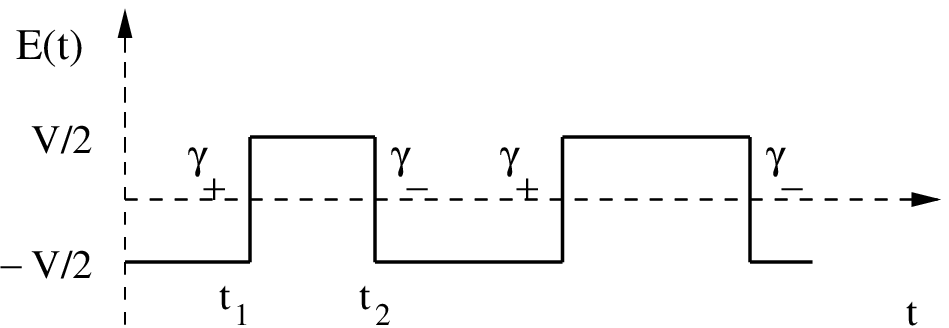}}}
\hspace{4mm}
\resizebox{!}{48mm}{\includegraphics{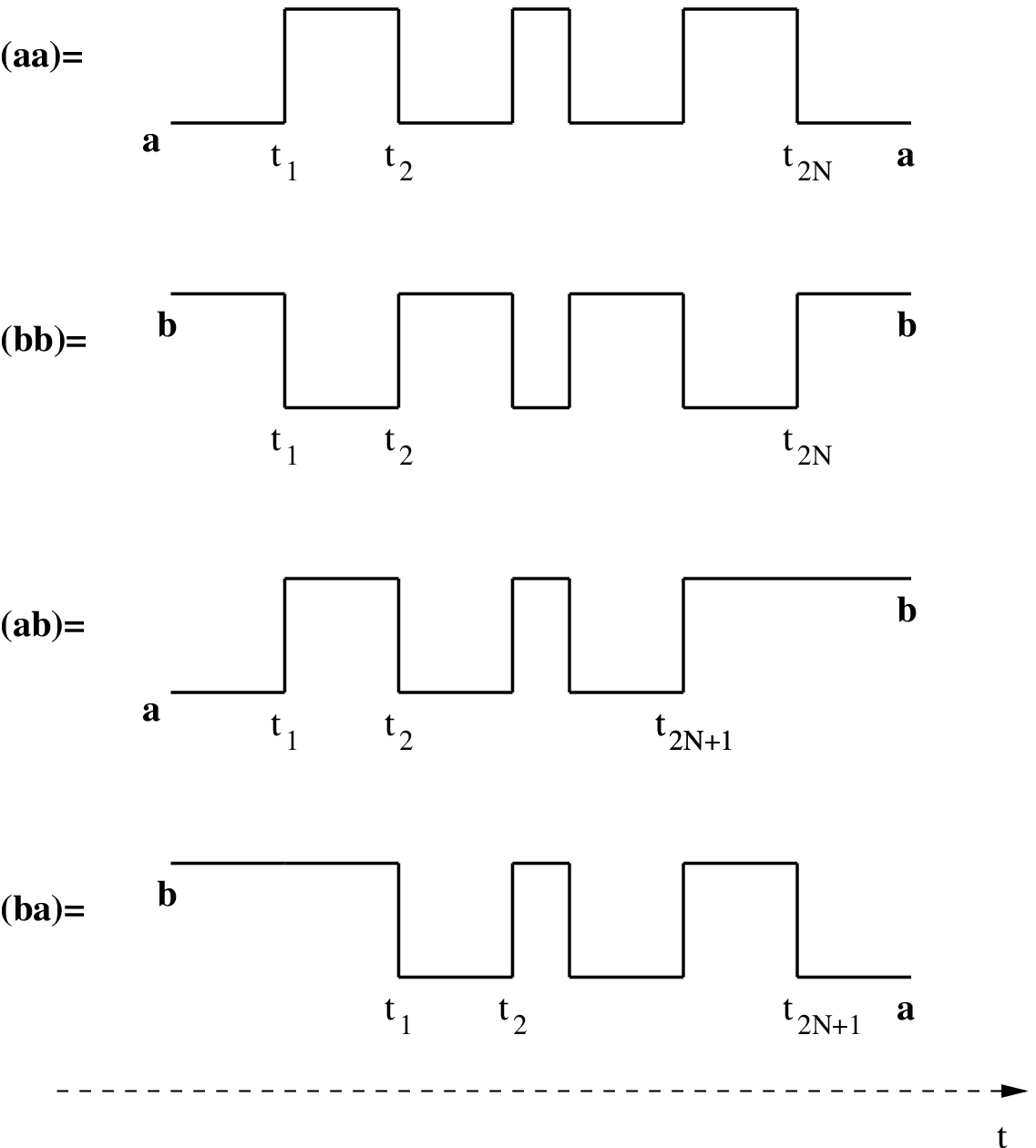}}
\end{center}
\caption{(a) Random Telegraph fluctuator switching between valor $\pm v/2$
 with rates $\gamma_{\pm}$;
(b) Each fluctuator can follow four kinds of paths. If intial and final states are equal, the path has an even number of transitions, an odd number of flips occurs if initial and final states are different.}
\label{rtn}
\end{figure}

We turn now to the derivation of Eq.(\ref{offdiag}) for an environment of 
classical fluctuators. 
Eq.(\ref{eq:semi2}) has to be averaged over the possible realizations of 
the stochastic process ${E}(t)$ with given initial conditions $\delta p_j^0$.
As already discussed we need to perform the calculation only for a  
single Random Telegraph process switching between the two values 
$E_a(t)= - v/2$ and $E_b(t)= v/2$ with rates 
$\gamma_{+}$ and $\gamma _-$. (see Fig.(\ref{rtn}a)).

The average $\overline{E(t)}$ is related to the average population 
difference between the two state $a,b$ of the BC,
$\overline{\delta p(t)}=\left ( \delta p^0 - \overline{\delta p} 
\right ) e^{-\gamma t} + \overline{\delta p}$, by 
$\overline{E(t)}=-v/2 \overline{\delta p(t)}$ , where 
$\gamma=\gamma_+ + \gamma_-$ is the total switching rate, 
$\delta p^0$ and 
$\overline{\delta p}=(\gamma_- - \gamma_+ )/\gamma$ are 
respectively the initial and equilibrium populaztion difference. 
For a stationary process $\delta p^0=\overline{\delta p}$ the Fourier 
transform 
of the second cumulant $\overline{[E(t)E(t+\tau)]}_c$ gives the power 
spectrum of the polarization fluctuations $ v^2/2 
(1- \overline{p}^2) \, \gamma/(\gamma^2 + \omega^2)$.

In order to calculate the generating functional 
$Z(t) \equiv \overline{ \exp \{-i \int_0^t dt' E(t') \}}$, we have to take 
into account all the possible realizations of the stochastic processes 
$E(t)$ with fixed initial conditions for the fluctuator, $\delta p^0$. 
To this end we have to consider that the fluctuator can follow one of the 
four kinds of paths illustrated in Fig.\ref{rtn}b. 

Then we have to perform (i) a sum over all possible number of switches, 
(ii) an integration over the corresponding transition times. 
For each process in Fig.\ref{rtn}b we can define a probability density 
$\pi_{\alpha,\beta}(t;t_m-t_1|0)$,($\alpha,\beta$ can be either $a$ or $b$) 
of finding the fluctuator in the state $\beta$ at time $t$ once it has 
been prepared in the state $\alpha$ at time $t=0$ after $m$ transitions 
at times $t_i$, $i=1,...m$. 
The probability densities of the elementary processes of having at most a 
single transition in a fixed time interval are easily found from the 
statistics of the populations of the two states $a$ and $b$,
$\overline{p}_{\alpha} (t)$, $\alpha=a,b$. 
The probability densities to stay in the states $a$ and $b$ are 
$\displaystyle{\pi_{aa}(t_2|t_1)=\exp[-\gamma_+(t_2-t_1)]}$, and 
$\displaystyle{\pi_{bb}(t_2|t_1)=\exp[-\gamma_-(t_2-t_1)]}$. 
The probability density of a single flip at time $t_1$ in the interval 
$t-t_0$ for instance from $a$ to $b$ reads 
$\displaystyle{\pi_{ab}(t;t_1|t_0)=\exp[-\gamma_-(t-t_1)] 
\gamma_+ \exp[-\gamma_+(t_1-t_0)]}$. 
With this, the probability density of the complex processes 
$\displaystyle{\pi_{\alpha,\beta}(t;t_m-t_1|0)}$ are given by:
\begin{eqnarray}
\pi_{aa}(t;t_{2N}-t_1|0)
&=& (\gamma_+ \gamma_-)^N e^{-\gamma_+ T_a - \gamma_- T_b}
\, = \,(\gamma_+ \gamma_-)^N e^{-\gamma_+t - \gamma_- \overline{\delta p} T_{2N}}\; 
	\nonumber \\
\qquad \,	\pi_{ab}(t;t_{2N+1}-t_1|0)&=&(\gamma_+ \gamma_-)^N \gamma_+ 
	e^{-\gamma_+ T_a - \gamma_- T_b}
\, = \, (\gamma_+ \gamma_-)^N \gamma_+ e^{-\gamma_-t - \gamma_+ 
	\overline{\delta p} T_{2N+1}}\;
\label{st1}
\end{eqnarray}
The probability densities $\displaystyle{\pi_{bb}(t;t_m-t_1|0)}$, 
$\displaystyle{\pi_{ba}(t;t_m-t_1|0)}$ are found from the expressions  
$\displaystyle{\pi_{aa}(t;t_m-t_1|0)}$ and 
$\displaystyle{\pi_{ba}(t;t_m-t_1|0)}$ with the replacements $\gamma_+ \to
\gamma_-$ and $\overline{\delta p} \to - \overline{\delta p}$. 
Each $\displaystyle{\pi_{\alpha,\beta}(t;t_m-t_1|0)}$ only depends on the 
number $m=2N, \, (2N+1)$ of switches and on the total times $T_{\alpha}$ 
during the random fluctuator has assumed value $E_{\alpha}$. Thus we have:
\begin{eqnarray}
&&
Z(t)
	= \sum_{\alpha,\beta} p_{\alpha}^0 z_{\alpha,\beta}(t) 
\;\nonumber \\ &&
z_{\alpha,\beta}(t)=\sum_{m=0}^{\infty} \int_0^t dt_m ... \int_0^{t_2} dt_1 
\pi_{\alpha,\beta}(t;t_m-t_1|0) e^{-i(E_\alpha T_\alpha + E_\beta T_\beta)}, \;
\end{eqnarray}
where $p_\alpha^0$ is the probability to find the fluctuator in the state 
$\alpha$ at time $t=0$, and $m=2N$ if $\alpha=\beta$ 
(the term $m=0$ being $1$), $m=2N+1$ if $\alpha \neq \beta$. 
The total times spent by the fluctuators in the state $a$ or $b$, $T_a$ and 
$T_b$, are given respectively by
\begin{eqnarray}
T_b&=& \sum_{n=1}^{2N}(-1)^n t_n \equiv T_{2N} \, 
	\qquad \qquad \qquad \qquad \quad
T_a=t-T_{2N} \qquad   \alpha=\beta \;\nonumber\\
T_b&=& \sum_{n=1}^{2N}(-1)^n t_n +t-t_{2N+1} \equiv t+ T_{2N+1} \, 
	\quad T_a=-T_{2N+1} \qquad \alpha \neq \beta \;
\label{st2}
\end{eqnarray}
From eqs. (\ref{st1}), (\ref{st2}) $Z(t)$ can be written in the simple form
\begin{equation}
Z(t)=A e^{-\frac{\gamma}{2}(1-\alpha)t}+(1-A) 
e^{\frac{\gamma}{2} (1+\alpha)t}\;
\label{an1}
\end{equation}
where $A$ and $\alpha$ are defined in Eq.~(\ref{alpha}). 
An interesting approximation of Eq.(\ref{an1}) is obtained by taking its 
second cumulant expansion. This amounts to consider a Gaussian process 
$E(t)$ having the same power spectrum $S(\omega)$.
Estimating the average given in 
Eq.(\ref{an1}) 
by its second cumulant and taking $\delta p^0=\overline{\delta p}$ we get
Eq.(\ref{eq:gamma-pure-dephasing-oscillators}).

The above exact results of the semiclassical analysis represent an important 
limiting case of decoherence due to a {\em quantum} discrete environment. 
Eq. (\ref{an1}) reproduces in fact the results 
presented in Section~\ref{nearly} for the pure dephasing decay due to 
an environment 
of quantum bistable fluctuators, in the regime in which each fluctuator 
performs a relaxation dynamics. \\

We thank L. Faoro, A. D'Arrigo and F. Taddei who 
collaborated to part of the work presented here. We warmly thank G. Giaquinta
who initiated the work on mesoscopic physics at the University of Catania. 
We thank M. Palma, A. Shnirman, G. Sch\"on and C. Urbina for discussions 
which greatly sharpened the point of view presented in this work. 
Very useful discussion with L. Amico, J. Clarke, D. Esteve, F. Hekking, 
P. Lafarge, J. Friedman, M. Grifoni, D. van Harlingen, J.E. Mooji, 
Y. Nakamura, Y. Nazarov, F. Plastina, J. M. Raimond, V. Tognetti, U. Weiss, 
D. Vion and A. Zorin are finally acknowledged.


\begin{thebibliography}{399}
\bibitem{kn:qcomp} 
	A. Ekert and A. Jozsa, Rev. Mod. Phys., {\bf 68}, {733} (1996);
 	{\em Quantum Computation and Quantum Information Theory},
	edited by 
	C. Macchiavello, G.M. Palma, A. Zeilinger,
	World Scientific (2000).
\bibitem{kn:nielsen}	
	{M.Nielsen and I.Chuang}
        \em{Quantum Computation and Quantum Communication},
        Cambridge University Press, (2000).
\bibitem{kn:clark-ed}	
 	\em{Experimental implementation of quantum computation},
	edited by {R.G. Clark}, Rinton Press, Princeton (2001).  
\bibitem{kn:loss} 
	{D. Loss and  P. Di Vincenzo}
	{Phys. Rev. A}, {\bf 57}, {120} (1998).
\bibitem{kn:rmp} 
	{Y. Makhlin, G. Sch\"on and A. Shnirman}, 
	{Rev. Mod. Phys.}, {\bf 73}, {357} (2001).
\bibitem{kn:teor}
	{D.A.\ Averin} {Sol.\ State Comm.}, {\bf 105}, {659} (1998);
	{L.B.\ Ioffe {\em et al.}} {Nature}, {\bf 398}, {679} (1999);
	{J.E.\ Mooij {\em et al.}} {Science}, {\bf 285}, {1036} (1999).
\bibitem{makhlin99} 
	{Y. Makhlin, G. Sch\"on and A. Shnirman}
	{Nature}{ \bf 398}, {305} (1999); 
	{A. Shnirman, G. Sch\"on and Z. Hermon} 
	{Phys.\ Rev.\ Lett.}{\bf 79}, {2371} (1997).
\bibitem{kn:nature} 
	{G. Falci \em et al.} {Nature}, {\bf 407}, {355} (2000).
\bibitem{kn:Nakamura1} 
	{Y.\ Nakamura, Yu.A.\ Pashkin, J.S.\ Tsai} 
	{Nature}, {\bf 398}, {786} (1999).
\bibitem{kn:vion} {D. Vion {\em et al.} %
} 
	{Science}{\bf 296}, {886} (2002); 
	{Y. Yu \em et al.} {Science}, {\bf 296}, {889} (2002);
	{J. Martinis \em et al.} {Phys. Rev. Lett.}, {\bf 89}, {117901} (2002);
	{J. Friedman \em et al.} {Nature}, {\bf 406}, {43} (2000);
	{I. Chiorescu {\em et al.}} private communication.
\bibitem{kn:two-qubit} {Yu. A. Pashkin {\em et al.}} cond-mat/0212314.
\bibitem{kn:zurek}
	{W.\ Zurek} {Physics Today}, {\bf 44}, {36} (1991).
\bibitem{kn:palma96}
	{G.M.Palma, K.-A.Suominen and A.K.Ekert}
	{Proc. Roy. Soc. London A}, {\bf 452}, {567} (1996).
\bibitem{kn:weiss} {U. Weiss} {\em Quantum Dissipative Systems} 2nd Ed 
	(World Scientific, Singapore 1999).               
\bibitem{kn:leggett} 
	{A. Leggett et. al} {Rev. Mod. Phys.}, {\bf 59}, {1} (1987).
\bibitem{kn:lax-redfield} 
	{A. G. Redfield} {IBM J. Research Develop}, {\bf 1}, {19} (1957);
	{M. Lax} {Phys. Rev.}, {\bf 145}, {110} (1966).
\bibitem{kn:cohen} 
	{C. Cohen-Tannoudji, J. Dupont-Roc and G. Grynberg}
	{\em Atom-Photon Interactions}, Wiley-Interscience (1993)
\bibitem{kn:PRL}
	{E.\ Paladino {\em et. al.}} 
	{Phys. Rev. Lett.}, {\bf 88}, {228304} (2002).
\bibitem{zorin96} 
	{A.B.\ Zorin {\em et. al}}
	{Phys.\ Rev.\ B}, {\bf 53}, {13682} (1996).
\bibitem{nakamura-echo} 
	{Y. \ Nakamura {\em et. al.}} 
	{ Phys. Rev. Lett.}, {\bf 88}, {047901} (2002).
\bibitem{kn:weissman} 
	{M.B.\ Weissman} {Rev.\ Mod.\ Phys.}, {\bf 60}, {537} (1988). 
\bibitem{kn:noise-set}
	{M.\ Covington {\em et al.}}
	{Phys.\ Rev.\ Lett.}, {\bf 84}, {5192} (2000).
\bibitem{bauernschmitt93} 
	{R.\ Bauernschmitt and Y.V.\ Nazarov} 
	{Phys.\ Rev.\ B}, {\bf 47}, {9997} (1993).
\bibitem{kn:mahan} {G.D. Mahan} {\em Many-Particle Physics}	
	Kluwer Academic, New York, (2000)
\bibitem{kn:shnirman} 
	{A. Shnirman,Y. Makhlin, G. Sch\"on}, Physica Scripta, {\bf T102},
	147, (2002).
\bibitem{kn:cottet}
	{A. Cottet {\em et al.}}
	in {\em Macroscopic Quantum Coherence and Quantum Computing}
	edited by {D.V. Averin, B. Ruggiero and P. Silvestrini},
	(Kluwer Pub., 2001),pg.111. 
\bibitem{kn:tian00}
	{L. Tian {\em et al.}}
	in {Proceedings of the NATO-ASI on Quantum Mesoscopic Phenomena 
	and Mesoscopic Devices in Microelectronics},
	edited by {I.O. Kulik and R. Elliatioglu} 
	(Kluwer Pub. 2000), pg. 429.
\bibitem{kn:mooji} {J.E.\ Mooij {\em et al.}}, 
	{Science}, {\bf 285}, {1036} (1999).
\bibitem{kn:clark} {D. J. Van Harlingen {\em et al.}} in 
	{\em Quantum Computing and Quantum Bits in Mesoscopic Systems}
	 Proceedings of the International Workshop on ``Macroscopic
	Quantum Coherence and Computing'', 
	Napoli 3-7 Giugno 2002  
	(Kluwer Academic Plenum Press)  (2003). 
\bibitem{kn:mqc2} {E. Paladino {\em et al.}}	
	in {\em Quantum Computing and Quantum Bits in Mesoscopic Systems}, 
	 Proceedings of the International Workshop on ``Macroscopic
	Quantum Coherence and Computing'', 
	Napoli 3-7 Giugno 2002  
	(Kluwer Academic Plenum Press) (2003). 
\bibitem{kn:plastina} {F. Plastina and G. Falci}, Phys. Rev. B, {\bf 67}, 
	224514 (2003)
\bibitem{kn:kouvenowen} {R. Aguado and L. P. Kouwenhowen} 
	{Phys. Rev. Lett.}{\bf 84}, {1986} (2000)
\bibitem{kn:dissipative} {A. O.Caldeira and A. J. Leggett}
	{Ann. Phys.}{\bf 149}, {374} (1983).
\bibitem{kn:LTcircuit} {E. Paladino {\em et. al}} {Physica E}
	{\bf 18}, 39 (2003)
\end{thebibliography}
\end{document}